\documentstyle[epsf]{article}

\font\ten=cmbx10 at 12pt

\newdimen\unit
\def\point#1 #2 #3{\vbox to0pt{\kern-#2\unit
  \hbox{\kern#1\unit#3}\vss}
 \nointerlineskip}
\def\plb#1{Phys.~Lett.~{\bf B#1}}
\def\npb#1{Nucl.~Phys.~{\bf B#1}}
\def\prl#1{Phys.~Rev.~Lett.~{\bf #1}}
\def\prd#1{Phys.~Rev.~{\bf D#1}}
\def\zpc#1{Z.~Phys.~{\bf C#1}}

\def\bpi{\bar B^0\to\pi^+\ell^-\bar\nu_\ell}
\def\brho{\bar B^0\to\rho^+\ell^-\bar\nu_\ell}
\def\bpigen{\bar B\to\pi\ell\bar\nu_\ell}
\def\brhogen{\bar B\to\rho\ell\bar\nu_\ell}

\def\expo#1{\,\times\,10^{#1}}

\def\g{\gamma}
\def\k{\kappa}
\def\a{\alpha}

\def\w{\omega}

\def\gm{\g^\mu}

\def\l{\left}
\def\r{\right}
\def\ord#1{{\cal O}\l(#1\r)}
\def\la{\langle}
\def\ra{\rangle}

\def\im{\mbox{Im}\,}
\def\ln#1{\mbox{ln}\l(#1\r)}

\def\tab#1{Table~\ref{#1}}

\def\fig#1{Fig.~\ref{#1}}

\def\sec#1{Section \ref{#1}}
\def\eq#1{Eq.~(\ref{#1})}
\def\eqs#1#2{Eqs.~(\ref{#1}) and (\ref{#2})}
\def\app#1{Appendix \ref{#1}}
\def\re#1{Ref.~\cite{#1}}

\def\lqcd{\Lambda_{QCD}}
\def\gbbpi{g_{B^*B\pi}}
\def\br#1{{\cal B}\l(#1\r)}
\def\vub{$V_{ub}$}

\epsfverbosetrue

\newcommand{\plus}{\makebox[15pt][c]{$+$}}
\newcommand{\minus}{\makebox[15pt][c]{$-$}}
\newcommand{\figurebox}[2]{\fbox{\vbox to#2in{\hbox to #1in{\hfil} \vfil}}}
\newcommand{\errr}[2]{\raisebox{0.08em}{\scriptsize 
{$\;\begin{array}{@{}l@{}}
                          \plus\makebox[0.9em][r]{#1} \\[-0.12em] 
                          \minus\makebox[0.9em][r]{#2} 
                        \end{array}$}}}

\newcommand{\er}[2]{\raisebox{0.08em}{\scriptsize {$\;\begin{array}{@{}l@{}}
                          \plus\makebox[0.15em][r]{#1} \\[-0.12em] 
                          \minus\makebox[0.15em][r]{#2} 
                        \end{array}$}}}

\newcommand{\be}{\begin{equation}}
\newcommand{\ee}{\end{equation}}
\newcommand{\bea}{\begin{eqnarray}}
\newcommand{\eea}{\end{eqnarray}}

\newcommand{\mev}{\,\mbox{MeV}}
\newcommand{\gev}{\,\mbox{GeV}}
\newcommand{\msbar}{\overline{\mbox{MS}}}

\newcommand{\bd}{\mbox{$\bar B\rightarrow D^{(*)}$ \,}}
\newcommand{\bks}{\mbox{$\bar B \rightarrow K^*\gamma$ \,}}

\begin{document}

\begin{titlepage}
  
\begin{center}
  
\renewcommand{\thefootnote}{\fnsymbol{footnote}}
  
{\ten Centre de Physique Th\'eorique\footnote{
Unit\'e Propre de Recherche 7061} - CNRS - Luminy, Case 907}

{\ten F-13288 Marseille Cedex 9 - France }
  
\vspace{1.2 cm}
  
{\huge\bf Lattice-Constrained Unitarity Bounds for 
$\bpi$ Decays}
  
%\vspace{0.3 cm}
\vspace{0.8 cm}

\setcounter{footnote}{0}
\renewcommand{\thefootnote}{\arabic{footnote}}
  
{\Large\bf Laurent Lellouch}\footnote{{\bf e-mail}: lellouch@cpt.univ-mrs.fr}
  
\vspace{1.5 cm}
  
{\bf Abstract}
\end{center}

Lattice results, kinematical constraints and QCD dispersion relations
are combined for the first time to derive model-independent bounds for
QCD form factors and corresponding rates. To take into account the
error bars on the lattice results we develop a general formalism which
ascribes well-defined statistical properties to these bounds. We
concentrate on $\bpi$ decays because of the relative simplicity of the
analytical behavior of the relevant polarization functions and because
of their immediate phenomenological relevance. To determine the range
of applicability of the dispersive matching required to obtain the
bounds, we have evaluated the leading perturbative and
non-perturbative QCD corrections to the relevant polarization
functions. Despite the lattice results' large error bars and limited
kinematical range, the bounds we obtain appear to favor certain
parametrizations for the form factors over others. The bounds also
enable the determination of $|V_{ub}|$ from the total experimental
rate with a theoretical error ranging from 27\% to 37\%, depending on
the assumptions made. The techniques developed here are, in fact,
quite general and are not limited to use with lattice results nor to
semileptonic $\bpi$ decays.

%\vspace{1,5 cm}

\vfill
  


\bigskip

\noindent Number of figures : 4 (in 6 separate PostScript files)
  
\bigskip
  
\noindent August 1996 (Revised)
  
\noindent Marseille preprint CPT-95/P.3236; hep-ph/9509358
  
%\noindent UGVA-DPT 1995/04-887
  
\bigskip
  
\noindent anonymous ftp or gopher: cpt.univ-mrs.fr
  
\end{titlepage}

\section{Introduction}
\label{intro}

The CLEO Collaboration has very recently reported preliminary
measurements of the total rates for semileptonic $\bpi$ and 
$\brho$ decays~\cite{bpicleo}. 
\footnote{Here and in the following, $\ell$ stands for the $e$ and 
the $\mu$ but not the $\tau$.} The study of these rates is important,
for it will eventually lead to accurate determinations of the
magnitude of the poorly known Cabibbo-Kobayashi-Maskawa (CKM) matrix
element \vub.
\footnote{The Particle Data Group reports $0.002\le |V_{ub}|\le 0.005$
~\cite{pdg}}
Such determinations, however, require precise
calculations of the relevant form factors which are
non-perturbative QCD functions of $q^2$, where $q$ is the
four-momentum transferred to the leptons.  These calculations are
difficult because they involve understanding the underlying QCD
dynamics over a large range of momentum transfers {}from
$q^2=q^2_{max}=(m_B-m_{\pi(\rho)})^2=26.4\gev^2(20.3\gev^2)$, 
where the final state hadron is
at rest in the frame of the $B$ meson, to $q^2\simeq 0$, where it
recoils very strongly. 

There exist several quark model determinations of these form factors.
The authors of \re{NaIS89} (ISGW) determine these form factors around
$q^2_{max}$ in a non-relativistic model and extrapolate
them in a somewhat {\it ad hoc} manner to the fully relativistic
region around $q^2=0$. The WSB model~\cite{WSB85} is a first attempt
at a relativistic, quark-model treatment of these decays, 
but only determines the form factors at $q^2=0$ and
extends them to arbitrary $q^2$ by assuming pole behavior. The
authors of \re{KS88} follow the WSB method but use pole or dipole
behavior according to the power-counting rules of QCD. Since these
rules do not modify the WSB results for $\bpi$ decays, we shall
not consider the results of \re{KS88} any further.
The authors of \re{PaOX95} also obtain the relevant 
form factors at $q^2=0$, 
in a light-front quark model.  In \re{DaSI95} these form factors
are calculated in the ISGW2 model which is a version of the 
ISGW model that is more consistent with Heavy Quark
Symmetry (HQS) constraints, more consistent with relativity and 
better behaved at
large recoils.  
None of these quark model calculations, however, determines the full
dynamics of the decay process. Only very recently have light-front
quark models begun to be used to determine the $q^2$-dependence of 
$\bar B\to \pi\ell\bar\nu_\ell$ decay form factors over a large
range of $q^2$~\cite{ChCHZ96,IGNS96}.

The form factors relevant to $\bar B\to\pi\ell\bar\nu_\ell$ decays 
have also been calculated using QCD
sumrules. Most of these calculations determine the form factors at
one or two values of $q^2$ and assume a $q^2$-dependence to obtain
the form factors at arbitrary $q^2$~\cite{CeDP88}-\cite{StN95}. In
\re{StN95}, though, this $q^2$-depedence is calculated to leading order in
$1/m_b$ and in Refs.~\cite{PaB93} and \cite{VlBK93} 
over a relatively large range of $q^2$ {}from
0 to about $15-20\gev^2$.  The authors of \re{VlBK93} even suggest,
in \re{VlBB95}, that their result for the form factor $f^+(q^2)$, which
dominates the rate for $\bar B\to\pi\ell\bar\nu_\ell$ decays, 
could be extended to $q^2_{max}$ by matching
it to a $B^*$ dominance model whose normalization is fixed with
the same sumrule. However, agreement between the various sumrule results
is not very good.

It is interesting to note that HQS, which is so useful for the study
of semileptonic \bd decays, where it provides a normalization
condition at zero recoil and applies to the full, physical,
kinematical range, is in principle valid for $heavy\to light$ quark
decays only around
$q^2=q^2_{max}$ and provides no normalization.  One can nevertheless
investigate the possibility of using HQS to relate $\bpigen$ decays to
$D\to\pi\ell\bar\nu_\ell$ decays, to obtain model-independent
information about the former~\cite{NaIW90}-\cite{LiW92}. This approach
is taken one step further in Ref.~\cite{GuBLNN94} where the structure
of the order $1/m_b$ corrections to the HQS limit description of
$\bpigen$ decays is explored.

Attempts have also been made to account for 
the short distance effects which become important at
large recoils by factorizing $\bpigen$ decays into a hard
perturbative part and a soft non-perturbative part, but it is unclear
whether the $b$ quark is massive enough for this factorization to
work~\cite{akhoury}.

On the lattice one is limited by cutoff effects. Present day
lattices have inverse lattice spacings of the order of $3\,\gev$ which
means that $B$ mesons cannot be simulated directly. One way of
preceeding is to perform the calculation for heavy quarks with masses
around that of the charm and extrapolate the results to $m_b$ using
heavy-quark scaling laws. One can also try to circumvent this problem
altogether by working with discrectized versions of effective 
theories such as Non-Relativistic QCD
(NRQCD) or Heavy-Quark Effective Theory (HQET) in which the mass of
the heavy quark is factored out of the dynamics. But even then the
momentum of the final state hadron in semileptonic $\bar B\to\pi$ and
$\bar B\to\rho$ decays is large compared to the cutoff in much of
phase space and, furthermore, these theories are only applicable around
$q^2_{max}$.  Thus, either approach will typically yield form factors
close to $q^2_{max}$ and one must perform extrapolations over a large
range of $q^2$ to reach small values of $q^2$ which contribute
significantly to the total rates.  The problem is even more accute for
\bks decays where one again obtains form factors close to $q^2_{max}$, 
$q$ being the photon momentum, while one needs them at the
on-shell point $q^2=0$.  The results of all these extrapolations will
depend very strongly on the assumptions made about the
$q^2$-dependence of the form factors.  Assuming nearest pole
dominance, ELC~\cite{ELC94} and APE~\cite{apebpi} have actually calculated the
rate for $\bpi$ decays {}from the determination of the relevant form
factor at a single $q^2\sim 18-20\gev^2$. However, their calculations provide
no information concerning the $q^2$-dependence 
of the form factor and therefore have large theoretical uncertainties.

An important step in constraining the $q^2$-dependence of form
factors in semileptonic 
$heavy\to light$ quark decays was taken by the UKQCD Collaboration
(D.~R.~Burford {\it et al.}) in
\re{jj} who use lattice results for the form factors for semileptonic
$\bpigen$ and \bks decays around $q^2_{max}$, kinematical constraints
at $q^2=0$ and heavy-quark scaling laws to select amongst different
possible functional forms. Though certain forms appear to be favored
in their analysis of semileptonic $\bpigen$ decays, the situation is
less clear for
\bks decays.  Furthermore, in the case of semileptonic $\brhogen$
decays, there are no kinematical constraints at $q^2=0$ for the form
factors that dominate the rate and their
analysis is not possible for these form factors. 

In light of all these limitations and discrepancies, a
fully model-independent QCD determination
of the $q^2$-dependence of the
form factors for semileptonic $heavy\to light$ decays is 
very important.  To this end we combine, in the present paper, 
QCD dispersion relations with the lattice results of the UKQCD
Collaboration~\cite{jj} and with a kinematical constraint and derive
model-independent bounds on the form factors relevant for $\bpi$ decays. 
The application to semileptonic $\bpigen$ decays of
the dispersive constraint techniques developed in
\re{machet} for semileptonic kaon decays was initiated
very recently by C.~G.~Boyd {\it et al.} in 
\re{boyd}\footnote{Dispersive bound techniques have also recently 
been applied to 
semileptonic $\bar B\to D^{(*)}\ell\bar\nu$~\cite{CGBGL96,IrCN96}
and $\Lambda_b\to\Lambda_c\ell\bar\nu$~\cite{CGBL96} decays.}.
Our analysis
extends theirs in many ways. The most notable difference is that
we use lattice results instead of model results to constrain the
bounds, thereby obtaining model-independent bounds which can be used
to test the consistency of experimental results or other
theoretical predictions with QCD itself.
\footnote{The use lattice results is briefly suggested 
in the conclusions of \re{boyd}.} 
The difficulty here is that one must develop a formalism to take into
account the errors on the lattice results. 

We also improve on the results of \re{boyd}
by obtaining bounds for the two form factors required to describe
$\bpigen$ decays--not only the dominant one--which enables us to make use
of a kinematical constraint to constrain
the form factors even further.  This again requires a generalization
of the bounding techniques of \re{machet}.

Moreover, we decompose the required polarization function according to
a more physical helicity basis.  This again enables us to improve on
the bounds of \re{boyd}. And unlike the analysis of \re{boyd}, ours
does not require knowledge of the $B^*B\pi$ coupling, $\gbbpi$. We in
fact use our results to put bounds on this coupling.  The only
physical parameters needed in our approach, then, are particle masses
and the leptonic $B^*$ decay constant, $f_{B^*}$, also available
{}from the lattice~\cite{qhldc}.

The remainder of the paper is organized as follows. In \sec{bkgnd} we
provide some general background, discuss the requirements of HQS and
describe the different parametrizations which we later compare with our
bounds. In \sec{method} we describe in some detail the methods we use
to obtain dispersive bounds on the form factors. In \sec{kin_cst}
we develop a formalism for taking into account kinematical constraints.
As a by-product, this formalism enables one to constrain bounds on
a form factor with the knowledge that it must lie within an interval
of values at one or more values of $q^2$. In \sec{errors} we develop
a formalism for taking into account uncertainties in results
used to constrain the bounds and obtain a probability which enables
us to define bounds with unambiguous statistical properties.
In \sec{bounds} we
combine all of the techniques of the previous sections with the lattice
results of \re{jj} to obtain
lattice constrained bounds for $f^0(q^2)$ and $f^+(q^2)$.
In \sec{comparison} we compare these
bounds to various parametrizations for the form
factors. In \sec{cplngbnds} we derive bounds and results for the
$B^*B\pi$ coupling. In
\sec{brbnds} we derive bounds for the total
rate for $\bpi$ decays and compare them to predictions
of other authors. We also compare some of these other authors'
predictions for $f^+(q^2)$ with our bounds on this form factor. 
In \sec{ccl} we summarize our main results and
discuss how they can be improved.  

In
\app{explicit}, we give explicit expressions for the bounds and
derive useful results. These can
easily be used, in conjunction with the other results in this paper,
to obtain model-independent bounds on form factors over
all of phase space from results obtained by any means in a limited
kinematical regime.  
In \app{pert} we investigate the range of
validity of the QCD calculation required to obtain the bounds.  We
find, for instance, that the choice of $Q^2=-q^2$ made in \re{boyd}, where
$q$ is the momentum flowing through the polarization function used to
obtain the bounds, leads to uncomfortably large perturbative and
non-perturbative corrections.
Finally, in \app{latpar} we present some of the parameters of the
lattice calculation and discuss systematic errors at length, including
large errors added to cover possible violations of flavor symmetry
in the light, active and spectator quarks. In
phenomenological applications and in situations where systematic errors
dominate over statistical errors--as they do here--we suggest that
they be taken into account earlier than they usually are in the
analysis of the lattice results, when one controls them better.  
Though one may loose some information in doing so, this
approach can also save one {}from drawing misleading conclusions.
Indeed, information about the $q^2$-dependence of the form factors
$f^+(q^2)$ and $f^0(q^2)$ may be lost if some of the systematic errors
taken into account, before fitting the form factors to different
functional forms, are independent of $q^2$.
\footnote{I would like to thank J.~Nieves for drawing my attention
to this point.} However, since the relative proportions of
$q^2$-independent and $q^2$-dependent systematic errors are not known,
we believe that it is safer to take into account all errors {}from the
beginning.

\section{General Background}
\label{bkgnd}

To describe semileptonic $\bpi$ decays one must evaluate the
matrix element
\be
\la \pi^+(p')|V^\mu|\bar B^0(p)\ra = 
\l(p+p'-q\frac{m_B^2-m_\pi^2}{q^2}\r)^\mu
f^+(q^2)+q^\mu\frac{m_B^2-m_\pi^2}{q^2}f^0(q^2)
\ ,
\label{matelt}
\ee
where $V^\mu{=}\bar u\gm b$ and $q=p-p'$. Here $q^2$ runs {}from
$m^2_{lepton}{\simeq}0$ to $q^2_{max}{=}(m_B-m_\pi)^2$.  $f^+$ and
$f^0$ correspond to the exchange of $1^-$ and $0^+$ particles,
respectively, and must satisfy the kinematical constraint at $q^2{=}0$,
\be
f^+(0)=f^0(0)
\ .
\label{kincst}
\ee
For $q^2\simeq q^2_{max}$, where HQS holds,
there are additional constraints. One can easily show that
HQS implies the following scaling of the form factors with $m_B$
at fixed $\w=(m_B^2+m_\pi^2-q^2)/(2m_Bm_\pi)$:
\be
f^0(q^2{\simeq}q^2_{max})= A^0(\w)\sqrt{\frac{1}{m_B}} 
\l(1+\ord{\frac{\lqcd}{m_B}}\r)
\label{hqsfo}
\ee
and 
\be
f^+(q^2{\simeq}q^2_{max})= A^+(\w) \sqrt{m_B} 
\l(1+\ord{\frac{\lqcd}{m_B}}\r)
\ ,
\label{hqsfp}
\ee
where $A^0(\w)$ and $A^+(\w)$ are independent of $m_B$.

Eqs.~(\ref{kincst}), (\ref{hqsfo}) and (\ref{hqsfp}) obviously limit the
relative functional forms of the two form factors. If we write
\be
f^0(q^2) = g(q^2)\, f^+(q^2)
\label{gdef}
\ee
then \eq{kincst} implies (as long as $f^+(0)\ne 0$)
\be
g(0)=1
\label{geqone}
\ee
and the heavy-quark scaling relations of \eqs{hqsfo}{hqsfp} imply
the following scaling relation for $g$:
\be
g(q^2{\simeq}q^2_{max}) = \frac{1}{m_B}\,\frac{A^0(\w)}{A^+(\w)}
\, \l(1+\ord{\frac{\lqcd}{m_B}}\r)
\ .
\label{gscaling}
\ee

Thus, in choosing parametrizations for $f^+$ and $f^0$ to compare
with our bounds we must 
ensure that the corresponding functions $g(q^2)$ satisfy 
\eq{geqone} and scale like $1/m_B$ as required by \eq{gscaling}.
One set of
parametrizations that is consistent with these requirements,
that is simple and physically motivated by pole
dominance ideas, is~\cite{jj}
\be
f^0(q^2)=f/(1-q^2/m_o^2)^n \mbox{\quad and\quad}
f^+(q^2)=f/(1-q^2/m_+^2)^{(n+1)}
\ , 
\label{npole}
\ee
where $n$ is fixed and $f=f^0(0)=f^+(0)$. $f$, $m_+$ and $m_o$ are the
parameters to be determined by a simultaneous fit to lattice results
for $f^+(q^2)$ and $f^0(q^2)$.  For the scaling relation of
\eq{gscaling} to be obeyed, $m_+$ and $m_o$ must both be equal to
$m_B$ up to $\ord{\lqcd/m_B}$ corrections. Moreover, because
$f^0(q^2)$ depends very weakly on $q^2$ in the range covered by the lattice
results we need
only consider the cases $n{=}0$ and $n{=}1$ which we shall call
``constant/pole'' and ``pole/dipole'' fits, respectively.

We will also compare our bounds on $f^+$ with the vector
dominance form
\be
f^+(q^2)=\frac{f^+(0)}{1-q^2/m_{B^*}^2}
\label{fppole}
\ ,\ee
which should be an accurate description of the form factor close to
zero recoil since the $B^*$ pole is so near
($m_{B^*}=5.32\gev$~\cite{pdg}).  Grinstein and Mende~\cite{BeGM94}
argue, in fact, that this behavior is likely to persist over the full
kinematical range, having first shown that it does in a combined
heavy-quark, chiral and large $N_c$ limit.  That this behavior
approximately persists is confirmed by Ball~\cite{PaB93} who actually
calculates the $q^2$-dependence of $f^+$ using three-point function
sumrules and also by Belyaev {\it et al.}~\cite{VlBK93} who obtain
this dependence with two-point function light-cone sumrules. In both
these calculations, however, the mass of the pole is slighly smaller
than $m_{B^*}$.  To find a $q^2$-dependence for
$f^0(q^2)$ which is consistent with the $q^2$-dependence of \eq{fppole} 
for $f^+(q^2)$ we note that in the heavy quark limit $f^-(q^2)=-f^+(q^2)$,
where $q^2\, f^-(q^2)=\l(m_B^2-m_\pi^2\r)\l(f^0(q^2)-f^+(q^2)\r)$.
This suggests that with $f^+(q^2)$ parametrized by \eq{fppole},
$f^0(q^2)$ may be well described by
\be
f^0(q^2)=(1-q^2/m_1^2)\frac{f^+(0)}{1-q^2/m_2^2}
\ ,\label{fzpole}
\ee
where $m_1$ and $m_2$ are
masses on the order of $m_B$. Note that \eqs{fppole}{fzpole} are consistent
with the kinematical and heavy-quark scaling constraints of 
\eqs{kincst}{gscaling}.
In what follows we refer to the combined
fit of $f^+(q^2)$ and $f^0(q^2)$ to the forms of \eqs{fppole}{fzpole} as the
``fixed-pole'' fit.

Finally, we will consider fits to \eqs{fppole}{fzpole} for different
values of ``$m_{B^*}$'' to determine the range of pole masses allowed by
our bounds.

\section{Dispersive Bounds}
\label{method}

The bounds on $f^+(q^2)$ and $f^0(q^2)$ 
are derived {}from the two-point function
\bea
\Pi^{\mu\nu}(q) &=& i\int d^4x\ e^{iq\cdot x} \la 0|T\l(V^\mu(x)
V^{\nu\dagger}(0)\r)|0\ra\nonumber\\
&=&-(g^{\mu\nu}q^2-q^\mu q^\nu)\,\Pi_T(q^2)+q^\mu q^\nu \,\Pi_L(q^2)
\ ,
\label{twopoint}
\eea
with $V^\mu=\bar u\g^\mu b$ and where $\Pi_{T(L)}(q^2)$ corresponds
to the propagation 
of a $J^P=1^-\,(0^+)$ particle. In the deep euclidean region, i.e.
$q^2=-Q^2$ with $Q^2\gg\lqcd$, this two-point function can reliably be
evaluated with perturbative QCD.  The extent to which the condition
$Q^2\gg\lqcd$ can be relaxed is investigated in \app{pert}. The
idea, then, is to use dispersion relations to relate this two-point
function to the matrix element of
\eq{matelt}. As long as the indices in $\Pi^{\mu\nu}(q)$ are treated
symmetrically the absorptive part, which is obtained by inserting a
complete set of hadronic states between the two factors of the 
vector current, will
be a sum of positive terms. Thus, the QCD result for this two point
function is an upper bound on the contribution {}from the $B\pi$
intermediate state. This bound enables us to constrain
$f^+(q^2)$ and $f^0(q^2)$.

Let us explore how this works in more detail.  The polarization
functions $\Pi_{T,L}(q^2)$ have a branch point at $q^2=(m_B+m_\pi)^2$
which marks the $B\pi$ threshold.  Below this threshold only one pole,
corresponding to the $B^*$ vector meson, is present. Because of its
quantum numbers this pole contributes to $\Pi_T(q^2)$ but not to
$\Pi_L(q^2)$. Thus, the spectral functions $\im\Pi_{T,L}(q^2)$, which
represent the absorptive parts of $\Pi_{T,L}(q^2)$ and which are
defined by
\bea
&& -(g^{\mu\nu}q^2-q^\mu q^\nu)\ \im\Pi_T(q^2)+q^\mu q^\nu\ \im\Pi_L(q^2)
\nonumber\\
&& =\frac{1}{2}\sum_\Gamma (2\pi)^4\delta^{(4)}\l(q-p_\Gamma\r)
\la 0|V^\mu|\Gamma\ra\la\Gamma|V^{\nu\dagger}|0\ra
\ ,
\label{specfn}
\eea
where the summation extends over all hadron states with the correct
quantum numbers, are given by 
\be
\im\Pi_L(t)\ge \frac{3}{2}\times\frac{t_+t_-}{16\pi}\sqrt{(t-t_+)(t-t_-)}
\frac{|f^0(t)|^2}{t^3}
\theta(t-t_+)
\label{pil}
\ee
and
\bea
\im\Pi_T(t) &\ge & \pi\l(\frac{m_{B^*}}{f_{B^*}}\r)^2\delta\l(t-m_{B^*}^2\r)
\nonumber\\
& + &\frac{3}{2}\times\frac{1}{48\pi}\frac{\l[(t-t_+)(t-t_-)\r]^{3/2}}{t^3}
|f^+(t)|^2 \,\theta(t-t_+)
\ ,
\label{pit}
\eea
where $t=q^2$ and $t_{\pm}=(m_B\pm m_\pi)^2$. In \eqs{pil}{pit} we have
limited ourselves to the contribution of the $B^*$ 
state with
\be
\la 0|V^\mu|B^*(r,p)\ra = \frac{m_{B^*}^2}{f_{B^*}}\epsilon^\mu_r
\label{bsdecaycst}
\ee
and to the contributions of intermediate $B\pi$ pairs. These latter
contributions appear with an overall factor of 3/2 because the two states
$|\bar B^0\pi^-\ra$ and $|B^-\pi^0\ra$ contribute in a ratio of 2 to 1 
to the spectral
functions 
in the limit of exact isospin symmetry. The inclusion
of other states would lead to stronger bounds on $f^+(q^2)$ and
$f^0(q^2)$ but would
require one to make assumptions concerning the size of their
contributions to the spectral functions.

Now, in QCD the polarization
functions $\Pi_{T,L}(q^2)$ satisfy the following dispersion
relations ($q^2=-Q^2$):
\bea
\chi_L(Q^2)=\l(-\frac{\partial}{\partial Q^2}\r) \l(-Q^2\Pi_L(Q^2)\r)=
\frac{1}{\pi}\int_0^\infty
dt\,\frac{t\,\im\Pi_L(t)}{\l(t+Q^2\r)^2}
\ ,
\label{disprelS}
\eea
and
\bea
\chi_T(Q^2)=\frac{1}{2}
\l(-\frac{\partial}{\partial Q^2}\r)^2 \l(-Q^2\Pi_T(Q^2)\r)=
\frac{1}{\pi}\int_0^\infty
dt\,\frac{t\,\im\Pi_T(t)}{\l(t+Q^2\r)^3}
\ .
\label{disprelV}
\eea
So, combining \eqs{disprelS}{disprelV}  with \eqs{pil}{pit}, we find
\be
\chi_L(Q^2)\ge \frac{1}{\pi}\int_{t_+}^\infty dt\ k^0_L(t,Q^2)|f^0(t)|^2
\label{chil}
\ee
and
\be
\chi_T(Q^2)\ge \l(\frac{m_{B^*}^2}{f_{B^*}}\r)^2
\frac{1}{\l[m_{B^*}^2+Q^2\r]^3}
+\frac{1}{\pi}\int_{t_+}^\infty dt\ k^+_T(t,Q^2)|f^+(t)|^2
\ ,
\label{chit}
\ee
where $k^0_L(t)$ and $k^+_T(t)$ can be obtained easily. 

To obtain the bounds, we proceed along the lines of \re{machet}.  Both
\eqs{chil}{chit} give us an upper bound on the weighted integral 
{}from $t=t_+$ to $\infty$ of the squared magnitude of a form factor.
Generically, we have
\be
J(Q^2)\ge\frac{1}{\pi}\int_{t_+}^\infty dt\ k(t,Q^2)|f(t)|^2
\ .
\label{jqsq}
\ee
To translate this information into a bound on the form factor $f(t)$
for $t$ in the range $[0,t_-]$,
we map the complex $t$-plane into the unit disc with the following
conformal transformation:
\be
\frac{1+z}{1-z}=\sqrt{\frac{t_+-t}{t_+-t_-}}
\ .
\label{map}
\ee
Then, the integral around the cut in \eq{jqsq} becomes an integral
around the unit circle and the physical region for semileptonic
$\bpigen$ decays (i.e. $t:\ 0\to t_-$) is mapped onto the segment of
the real line $z:\ (1>z_o>0)\to 0$, where $z_o=z(t{=}0)$.
Thus, \eq{jqsq} becomes
\be
J(Q^2)\ge\int_{|z|=1} \frac{dz}{2\pi i\, z} |\phi(z,Q^2)\,f(z)|^2
\ ,
\label{jsqz}
\ee
where we have used the fact that $k(t,Q^2)$ is a positive definite quantity.
Here and in what follows we use the freedom we have in defining
$\phi(z,Q^2)$ to make it real and positive.

Then we define an inner product on the unit circle via
\be
\la g|h\ra = \int_{|z|=1} \frac{dz}{2\pi i\, z} \bar g(z)\,h(z)
\label{innerprod}
\ee
so that the inequality of \eq{jqsq} can be written
\be
J(Q^2)\ge\la \phi f|\phi f\ra
\ .
\label{jqsqinner}
\ee
The value of the form factor at any point $z(t)$ within the unit circle 
can be obtained by considering the inner product of $|\phi  f\ra$
with the state $|g_t\ra$ such that
\be
g_t(z)=\frac{1}{1-\bar z(t)z}
\ .
\label{gt}
\ee
If $f(t)$ has no poles below the cut, then Cauchy's theorem yields
\be
\la g_t|\phi  f\ra = \phi\l(z(t),Q^2\r)\,f\l(z(t)\r)
\ .
\label{fztnopole}
\ee
Now, because of the positivity of the inner product, we know that
\be
\mbox{det}\l(
\begin{array}{cc}
\la \phi f|\phi f\ra & \la\phi f|g_t\ra\\
\la g_t|\phi f\ra & \la g_t|g_t\ra
\end{array}
\r)\ge 0
\ .
\ee
One can easily show that this condition on the determinant, together with
\eq{jqsqinner}, 
implies
\be
|f(t)|^2\le J(Q^2)\frac{1}{1-\l|z(t)\r|^2}
\frac{1}{\l|\phi\l(z(t),Q^2\r)\r|^2}
\ ,
\label{fboundsnoinfo}
\ee
which is the bound we are after. 
The beauty of the method of \re{machet} is that it enables one
to incorporate information about the form factor $f(t)$
to make the bounds more constraining. Suppose, for example, 
that we know the value of $f(t)$ at a discrete set of points $t_i$, 
$i=1\cdots N$.
Then, as above, the positivity of the inner product guarantees that
the determinant of the $(N+2)\times(N+2)$ matrix
\be
M(f(t),\vec f) = 
\l(
\begin{array}{ccccc}
\la \phi f|\phi f\ra & \la\phi f|g_t\ra & \la\phi f|g_{t_1}\ra &\cdots &
\la\phi f|g_{t_N}\ra \\
\la g_t|\phi f\ra & \la g_t|g_t\ra & \la g_t|g_{t_1}\ra&\cdots
& \la g_t|g_{t_N}\ra\\
\la g_{t_1}|\phi f\ra & \la g_{t_1}|g_t\ra & \la g_{t_1}|g_{t_1}\ra&\cdots
& \la g_{t_1}|g_{t_N}\ra\\
\vdots&\vdots&\vdots&\vdots&\vdots\\
\la g_{t_N}|\phi f\ra & \la g_{t_N}|g_t\ra & \la g_{t_N}|g_{t_1}\ra&\cdots
& \la g_{t_N}|g_{t_N}\ra
\end{array}
\r)
\label{matrix}
\ee
is positive semi-definite:
\be
\det{M(f(t),\vec f)}\ge 0
\ ,
\label{bndsineq}
\ee
where $\vec f = \l(f(t_1),\cdots,f(t_N)\r)$. This condition on the
determinant, together with \eq{jqsqinner}, leads to an inequality for
a quadratic polynomial in $f(t)$ which in turn leads to the more
restrictive bounds
\be
F_{lo}(t|\vec f)\le f(t)\le F_{up}(t|\vec f)
\ ,
\label{fbounds}
\ee
where the general expressions for $F_{lo}(t|\vec f)$ and $F_{up}(t|\vec f)$ 
are given 
in \app{explicit}.

\bigskip

Now, if $f(t)$ has a pole at $t=t_p$ away {}from the 
cut, \eq{fztnopole} becomes
\be
\la g_t|\phi  f\ra = \phi\l(z(t),Q^2\r)\,f\l(z(t)\r)+
\frac{\mbox{Res}\l(\phi f,z(t_p)\r)}{z(t_p)-z(t)}
\ ,
\label{fztpole}
\ee
since a simple pole in $f(t)$ at $t=t_p$ translates into a simple pole
in $f(z)$ at $z=z(t_p)$.\footnote{Generalization to the case where
$f(t)$ has more than one pole is straighforward.} The rest of the
argument leading to \eq{fbounds} remains unchanged and we have,
instead of
\eq{fbounds},
\be
\l(F_{lo}(t|\vec f)-\frac{1}{\phi\l(z(t),Q^2\r)}
\frac{\mbox{Res}\l(\phi f,z(t_p)\r)}{z(t_p)-z(t)}\r)
\le f(t)\le 
\l(F_{up}(t|\vec f)-\frac{1}{\phi\l(z(t),Q^2\r)}
\frac{\mbox{Res}\l(\phi f,z(t_p)\r)}{z(t_p)-z(t)}\r)
\ ,
\label{fboundspole}
\ee
where one must not forget the pole contributions to 
$\la g_{t_i}|\phi  f\ra$ in the matrix which determines 
$F_{lo}(t|\vec f)$ and $F_{up}(t|\vec f)$.

If one does not know the residue of $f(t)$'s pole at $t=t_p$, one
can still obtain bounds on $f(t)$ {}from the knowledge of the
position of the pole alone~\cite{josep}. All one has to do is perform
the replacement
\be
\phi(z,Q^2)\to \phi_p(z,Q^2)\equiv \phi(z,Q^2) 
\frac{z-z(t_p)}{1-\bar z(t_p)\,z}
\ ,
\label{phipdef}
\ee
where $t_p$ is assumed to be in the range $[t_-,t_+]$ so that $\phi_p$
is positive for $z=z(t)$ with $t\in [0,t_-]$.  Then, $\phi_p(z)\,f(z)$
does not have a pole at $z=z(t_p)$ so that $\la g_i|\phi_p f\ra$ is
given by \eq{fztnopole} with $\phi(z)$ replaced by $\phi_p(z)$.
Furthermore, because $(z-z(t_p))/(1-\bar z(t_p)\,z)$ has magnitude one
on the unit circle, $\la \phi_p f|\phi_p f\ra$ is equal to $\la \phi
f|\phi f\ra$ and the crucial QCD constraint of \eq{jqsqinner} is left
unchanged. Thus, the arguments which lead to the bounds of
\eq{fbounds} can be applied here and we find
\be
F_{lo}^p(t|\vec f)\le f(t)\le 
F_{up}^p(t|\vec f)
\ ,
\label{fboundsp}
\ee
where $F_{lo}^p(t|\vec f)$ and $F_{up}^p(t|\vec f)$ are the functions
$F_{lo}(t|\vec f)$ and $F_{up}(t|\vec f)$ of \app{explicit} obtained
by replacing $\phi$ with $\phi_p$.  Because the bounds obtained with
$\phi_p$ assume no knowledge of the residue of $f(t)$'s pole at
$t=t_p$, they will be looser than bounds which do assume some
knowledge of this residue.

Finally, the bounds of \eq{fboundsp} can be used to constrain
the residue $f_p$ of the pole of the form factor $f(t)$ at $t=t_p$. All
one has to do is multiply the inequalities of \eq{fboundsp} by $1-t/t_p$
and take the limit $t\to t_p^-$ of the resulting functions.
With the definition
\be 
f(t) = \frac{f_p}{1-t/t_p}+f_{reg}(t)
\ ,
\label{ftpole}
\ee
where $f_{reg}(t)$ is finite at $t=t_p$, we find
\be
f^p_{lo}\le f_p\le f^p_{up}
\ ,
\label{resbndone}
\ee
with
\be
f_{lo(up)}^p = 4\frac{t_+-t_-}{t_p}\l(\frac{1+z(t_p)}{1-z(t_p)}\r)^2
F_{lo(up)}^p(t_p|\vec f)
\ .
\label{resbndtwo}
\ee
\section{Implementing the Kinematical Constraint}
\label{kin_cst}

\subsection{The Kinematical Constraint}
In \sec{method} we describe how to obtain independent bounds on the 
form factors
$f^+(t)$ and $f^0(t)$ given the values of these form factors at discrete
sets of points. The two form factors, however, are not entirely independent
since the kinematical constraint of \eq{kincst} requires that
$f^+(0)=f^0(0)$. In the present section we describe how to incorporate
this constraint into bounds for $f^+(t)$ and $f^0(t)$.

Without this kinematical constraint but given $f^+(t)$ at $t=t_1^+,\cdots,
t_{N^+}^+$ ($\vec f^+=(f^+(t_1^+),\cdots,f^+(t_{N^+}^+))$) and 
$f^0(t)$ at $t=t_1^0,\ldots,
t_{N^0}^0$ ($\vec f^0=(f^0(t_1^0),\cdots,f^0(t_{N^0}^0))$), the 
methods of \sec{method} yield:
\footnote{Here, $F$ stands for either $F$ and $F^p$.}
\be
F^+_{lo}(t|\vec f^+)\le f^+(t)\le F^+_{up}(t|\vec f^+)
\label{bndfp}
\ee
and
\be
F^0_{lo}(t|\vec f^0)\le f^0(t)\le F^0_{up}(t|\vec f^0)
\label{bndfz}
\ .\ee

Combined with the kinematical constraint, these bounds yield the following
for $f_o\equiv f^+(0)=f^0(0)$:
\footnote{We assume here that \protect\eqs{bndfp}{bndfz} are consistent
with the kinematical constraint at $t=0$ and will leave to 
\protect\sec{errors} the discussion of what happens if they are not.}
\be
\phi_{lo}\le f_o\le \phi_{up}
\label{fobnds}
\ ,\ee
with
\be
\phi_{lo}\equiv \mbox{Max}\l(F^+_{lo}(t|\vec f^+),F^0_{lo}(t|\vec f^0)\r)
\label{philodef}
\ee
and
\be
\phi_{up}\equiv \mbox{Min}\l(F^+_{up}(t|\vec f^+),F^0_{up}(t|\vec f^0)\r)
\label{phiupdef}
\ .\ee
This is all that we know since the bounds of \eqs{bndfp}{bndfz} 
carry no indication as to 
the probability that $f^+(t)$ or $f^0(t)$ take any particular value 
within them.
Thus, implementing the kinematical constraint reduces to finding the bounds
on $f^+(t)$ and $f^0(t)$ given $\vec f^+$, $\vec f^0$ {\it and}
$f_o\in[\phi_{lo},\phi_{up}]$.

\subsection{Bounds on a Form Factor Given that It Lies within an
Interval at a Discrete Set of Points}

To find the bounds on a form factor $f(t)$ given $\vec f=(f(t_1),\cdots,
f(t_N))$ and $F_{lo}(t_{N+1}|\vec f)\le\phi_{lo}\le 
x\equiv f(t_{N+1})\le\phi_{up}
\le F_{up}(t_{N+1}|\vec f)$ (for the case at hand, $t_{N+1}=0$),
\footnote{We choose $[\phi_{lo},\phi_{up}]\subseteq [F_{lo}(t_{N+1}|\vec f),
F_{up}(t_{N+1}|\vec f)]$ 
since, as shown in \protect\app{explicit}, 
$x$ cannot lie outside $[F_{lo}(t_{N+1}|\vec f),
F_{up}(t_{N+1}|\vec f)]$.}
consider
the bounds $F_{lo}(t|\vec f,x)$ and $F_{up}(t|\vec f,x)$ 
obtained from the constraint
$\mbox{det}M(f(t),\vec f,x)\ge 0$ using the methods of \sec{method}.

According to \app{explicit}, $F_{up}(t|\vec f,x)$ increases
monotically with $x$, for fixed $t$ and $\vec f$, 
from $x=F_{lo}(t_{N+1}|\vec f)$ to $x=x_{max}$ where it
reaches its maximum value $F_{up}(t|\vec f,x_{max})=F_{up}(t|\vec f)$
and then decreases monotically until $x=F_{up}(t_{N+1}|\vec f)$ beyond
which it is not defined.  Similarly, $F_{lo}(t|\vec f,x)$ decreases
monotically from $x=F_{lo}(t_{N+1}|\vec f)$ to $x=x_{min}$ where it
reaches its minimum value $F_{lo}(t|\vec f,x_{min})=F_{lo}(t|\vec f)$
and then increases monotically until $x=F_{up}(t_{N+1}|\vec f)$ beyond
which it is not defined.

Therefore, the upper bound on $f(t)$ given $\vec f$ and
$x\in[\phi_{lo},\phi_{up}]$ may be defined as the maximum of
$F_{up}(t|\vec f,x)$ w.r.t. $x$ at fixed $t$. Thus,
\be
F_{up}(t|\vec f,x\in[\phi_{lo},\phi_{up}])=
\l\{\begin{array}{ll}
F_{up}(t|\vec f,\phi_{up}) & \mbox{ if $x_{max}>\phi_{up}$}\\
F_{up}(t|\vec f) & \mbox{ if $x_{max}\in[\phi_{lo},\phi_{up}]$}\\
F_{up}(t|\vec f,\phi_{lo}) & \mbox{ if $x_{max}<\phi_{lo}$}
\end{array}\r.
\ .\label{upbndkin}
\ee
Similarly, the lower bound may be defined as the minimun of
$F_{lo}(t|\vec f,x)$ w.r.t. $x$:
\be
F_{lo}(t|\vec f,x\in[\phi_{lo},\phi_{up}])=
\l\{\begin{array}{ll}
F_{lo}(t|\vec f,\phi_{up}) & \mbox{ if $x_{min}>\phi_{up}$}\\
F_{lo}(t|\vec f) & \mbox{ if $x_{min}\in[\phi_{lo},\phi_{up}]$}\\
F_{lo}(t|\vec f,\phi_{lo}) & \mbox{ if $x_{min}<\phi_{lo}$}
\end{array}\r.
\label{lobndkin}
\ .\ee
Graphically, these definitions correspond to defining the upper and
lower bounds as the upper and lower boundaries of the envelope
obtained by considering the set of pairs of bounds $(F_{lo}(t|\vec f,x),
F_{up}(t|\vec f,x))$ and
allowing $x$ to vary freely in the interval $[\phi_{lo},\phi_{up}]$.

Furthermore, the behavior in $x$ of the bounds $F_{lo}(t|\vec f,x)$ 
and $F_{up}(t|\vec f,x)$ implies that constraining $x$ to lie
within the interval $[\phi_{lo},\phi_{up}]$ can only increase the
strength of the bounds, i.e. the kinematical constraint will lead to bounds
that are at least as constrained as those obtained without it. In practice,
we will find that the kinematical constraint significantly improves
the bounds on $f^+(t)$, especially for small $t$.

Finally, the conclusions of \app{explicit} can be used to generalize
the above results to the situation where bounds on a form factor
$f(t)$ are sought, given $\vec f=(f(t_1),\cdots, f(t_N))$ and the
constraints $\phi^i_{lo}\le f(t_i)\le\phi^i_{up}$, $i=N+1,\cdots,N+M$,
for any $M$.

\section{Taking Errors into Account: Bounds and Probabilities}
\label{errors}

Up until now, we have assumed that the values $\vec
f=(f(t_1),\cdots,f(t_N))$ used as input to constrain the bounds on
form factor $f(t)$ were both exact and consistent with the dispersive
constraints. In the present section we show how the methods of \sec{method}
and \sec{kin_cst} have to be extended to deal with the more realistic
situation in which the input vector $\vec f$ has errors, as is the case
when it is given by the lattice, experiment or any other approximate
means. Thus, in what follows, we will assume that $\vec f$ is distributed
according to a normalized probability distribution ${\cal P}_{in}(\vec f)
\,[d^Nf]$.
When considering the two form factors $f^+(t)$ and $f^0(t)$, we will assume
that $\vec f^+=(f^+(t_1^+),\cdots,f^+(t_{N^+}^+))$ and 
$\vec f^0=(f^0(t_1^0),\cdots,f^0(t_{N^0}^0))$ are distributed according
to ${\cal P}_{in}(\vec f^+,f^0)\,[d^{N^+}f][d^{N^0}f]$.

The problem which arises when $\vec f$ is obtained in an approximation
to QCD, from a model or from experiment is that the inequality
of \eq{bndsineq} may not have a solution. In \app{explicit} we show that a
solution exists iff $\mbox{det} M(\vec f)\ge 0$, where $M(\vec
f)$ is the matrix of \eq{matrix} with the second row and column deleted, as
suggested by our notation. That is, a solution exists iff $\vec
f$ is itself consistent with the dispersive constraint. The measure
for $\vec f$ which ensures such consistency is:
\be
d\mu(\vec f) = [d^Nf]\,\theta\l(\mbox{det} M(\vec f)\r)
\ ,\ee
where $\theta(x)$ is the standard ``theta-funtion''.

When considering $\bar B\to\pi\ell\bar\nu$ decays, we must also ensure
that the kinematical constraint of \eq{kincst} is satisfied.
The measure for $\vec f^+$ and $\vec f^0$ which incorporates
the requirements that $\vec f^+$ and $\vec f^0$ be consistent
with the dispersive and kinematical constraints, is:
\bea
d\mu(\vec f^+;\vec f^0)& \equiv & [d^{N^+}f^+]\;[d^{N^0}f^0]\,
\theta\l(\mbox{det} M^+(\vec f^+)\r)\,
\theta\l(\mbox{det} M^0(\vec f^0)\r)\nonumber\\
& &\theta\l(F_{up}^0(0|\vec f^0)-F_{lo}^+(0|\vec f^+)\r)\,
\theta\l(F_{up}^+(0|\vec f^+)-F_{lo}^0(0|\vec f^0)\r)
\label{measure}
\ ,\eea
where $M^+(\vec f^+)$ and $M^0(\vec f^0)$ are the matrix of 
\eq{matrix} relevant for bounds on $f^+(t)$ and $f^0(t)$,
respectively.

\bigskip

We can use this measure and the probability distribution of the
input, ${\cal P}_{in}(\vec f^+,\vec f^0)\,[d^{N^+}f][d^{N^0}f]$, to 
define bounds which have a clear statistical meaning. In doing so, it is
important to keep in mind that, for fixed $\vec f^+$ and $\vec f^0$, 
the bounds given by the methods of \sec{method} come in pairs and 
carry with them no indication as to 
the probability that $f^+(t)$ or $f^0(t)$ take any particular value 
within them.

A probability which takes these two points into account is the
probability that a pair of bounds $\l(F^+_{lo}(t|\r.$ $\vec
f^+, f^+(0) \in$ $[\phi_{lo},${}$\phi_{up}]),$ $F^+_{up}(t|$ $\vec
f^+, f^+(0) \in$ $[\phi_{lo},$ $\l.\phi_{up}])\r)$ for $f^+(t)$, defined from
\eqs{upbndkin}{lobndkin} with $\phi_{lo}$ and $\phi_{up}$ given in
\eqs{philodef}{phiupdef}, lie within the interval
$[f^+_{lo},f^+_{up}]$ at a given $t$. This probability is
given by
\bea
{\cal P}_+(t;[f_{lo}^+,f_{up}^+]|\vec f^+;\vec f^0) 
\equiv \frac{1}{Z(\vec f^+;\vec f^0) }
\int & d\mu(\vec f^+;\vec f^0)\,\,{\cal P}_{in}(\vec f^+,\vec f^0)\,\,
\theta\l(f^+_{up}-F_{up}^+(t|\vec f^+,f^+(0)\in[\phi_{lo},\phi_{up}])\r)
\nonumber\\
&\times\,
\theta\l(F_{lo}^+(t|\vec f^+,f^+(0)\in[\phi_{lo},\phi_{up}])-f^+_{lo}\r)
\label{probp}
\ ,\eea
where 
\be
Z(\vec f^+;\vec f^0)\equiv \int 
d\mu(\vec f^+;\vec f^0)\,{\cal P}_{in}(\vec f^+,\vec f^0)
\ee
is the probability that $\vec f^+$ and $\vec f^0$ are consistent
with the dispersive and kinematical constraints and where
$d\mu(\vec f^+;\vec f^0)$ is given by \eq{measure}. 

Because only pairs of bounds which lie entirely within $[f^+_{lo},f^+_{up}]$ 
are counted, the probability of \eq{probp} is the minimum probability
that the form factor $f^+(t)$ take a value inside $[f^+_{lo},f^+_{up}]$ 
at $t$. A similar probability can be defined for the bounds on $f^0(t)$
and it is these two probabilities that we use in \sec{bounds} to 
plot various ``confidence level'' (CL) bounds for $f^+(t)$ and $f^0(t)$.

\section{Bounds on $f^+(q^2)$ and $f^0(q^2)$}
\label{bounds}

We may now derive bounds for $f^+(q^2)$ and $f^0(q^2)$.  We
proceed in stages to show how the various elements that we combine
contribute to improving the bounds. Thus, we begin by determining bounds
on $f^+(q^2)$ and $f^0(q^2)$ without imposing the kinematical
constraint of \eq{kincst} nor using lattice results. We then impose
the kinematical constraint and in a third step only, combine
the bounds with the lattice results of the UKQCD Collaboration~\cite{jj}.

As described in
\sec{method}, we use the inequality of
\eq{chil} to obtain bounds on $f^0(q^2)$. Because the 
$B_0^*(J^P=0^+)$ lies above the $B\pi$ threshold, its
contribution to $\im\Pi_L(t)$ and $f^0$ does not enter the
determination of the bounds. Thus, without the kinematical 
constraint nor the lattice results, the bounds
are given by \eq{fboundsnoinfo} 
with
\be
\phi(z,Q^2)=\sqrt{\frac{3t_+t_-}{4\pi}}\frac{1}{t_+-t_-}
\frac{1+z}{(1-z)^{5/2}}\l(\beta(0)+\frac{1+z}{1-z}\r)^{-2}
\l(\beta(-Q^2)+\frac{1+z}{1-z}\r)^{-2}
\ ,
\label{phiz}
\ee
where we have used the results of \re{okubo} and where
$\beta(t)=\sqrt{(t_+-t)/(t_+-t_-)}$. 

The bounds on $f^+(q^2)$ are obtained from \eq{chit}. Combining \eq{chit}
with \eq{jqsq} gives
\be
J^+(Q^2)\ge\frac{1}{\pi}\int_{t_+}^\infty dt\ k^+_T(t,Q^2)|f^+(t)|^2
\ ,
\label{jqsqpbnd}
\ee
where
\be
J^+(Q^2)=\chi_T(Q^2)-\l(\frac{m_{B^*}^2}{f_{B^*}}\r)^2
\frac{1}{\l[m_{B^*}^2+Q^2\r]^3}
%-\frac{1}{\pi}\int_{t_+}^\infty dt\, k^0_T(t,Q^2)|f^0(t)|^2
\ .
\label{jqsqp}
\ee
We take $f_{B^*}=33\er{4}{4}$
in \eq{jqsqpbnd} as given by the lattice in \re{qhldc} on the same
configurations as those used to obtain $f^+(q^2)$ and $f^0(q^2)$.
\footnote{The error was obtained by combining in quadrature the 
statistical error with an additional 10\% systematic error.}

Now, unlike $f^0(q^2)$, $f^+(q^2)$ has a pole at $q^2=m_{B^*}^2<t_+$.
This pole is due to the fact that the virtual $W$-boson emitted in
$\bpi$ decays can couple to the $\bar B\pi$ vertex through a $B^*$. At
leading order in
a Heavy-Meson Chiral Lagrangian formalism, the $B^*B\pi$ coupling is
of the form~\cite{MaW92}
\be
{\cal L}_{B^*B\pi}=g\,\mbox{Tr}\l\{B_a\bar B_bA^\mu_{ba}\g_\mu\g_5\r\}
\ ,
\ee
where $B_a$ is a field which annihilates pseudoscalar and vector
mesons composed of a $b$ quark and a light antiquark of flavor $a$;
$\bar B_a$ is the dirac conjugate of $B_a$; and
$A^\mu=-\partial^\mu\pi/f_\pi+\cdots$ with $f_{\pi^\pm}=131\mev$ (for
details, see for example \re{eric}). A similar interaction can be
written down for mesons composed of $\bar b$ antiquarks and light
quarks.  The contribution of this coupling is straighforward to
evaluate. We find that close to $q^2_{max}$ it contributes to $f^+(q^2)$ a
pole of the form of \eq{ftpole} with $q^2_{pole}=m_{B^*}^2$ and residue
\be
f_p = \frac{\gamma_{B^*B\pi}\,g\,\bar m}{\sqrt{2}f_\pi}
\frac{1}{f_{B^*}}=g_{B^*B\pi}
\frac{1}{2 f_{B^*}}
\ ,
\label{fp}
\ee
where $\bar m=\sqrt{m_{B^*}m_B}$ and where $\gamma_{B^{*+}B^+\pi^0}=-1$ and 
$\gamma_{B^{*+}B^0\pi^+}=\sqrt{2}$.

Since we do not know the value of the coupling $g$ and hence the
residue of $f^+(q^2)$ at $q^2=m_{B^*}^2$, the bounds that do not take into
account the kinematical constraint of \eq{kincst} nor the results
from the lattice are given by \eq{fboundsnoinfo} but with the
replacement of \eq{phipdef} and with
\be
\phi(z,Q^2)=\sqrt{\frac{1}{\pi\,(t_+-t_-)}}
\frac{(1+z)^2}{(1-z)^{9/2}}\l(\beta(0)+\frac{1+z}{1-z}\r)^{-2}
\l(\beta(-Q^2)+\frac{1+z}{1-z}\r)^{-3}
\ .
\label{phip}
\ee 
$\beta(t)$ is defined after \eq{phiz}. 

\medskip

For both $f^0(q^2)$ and $f^+(q^2)$, the QCD evaluation of the
corresponding subtracted polarization function ($\chi_L(Q^2)$ and
$\chi_T(Q^2)$, respectively) is performed in \app{pert}.  Because this
evaluation appears to break down when $Q^2=-16\gev^2$--the value used
in~\re{boyd}--and because the choice of $Q^2$ does not change the
lattice constrained bounds that we derive below significantly, we take
$Q^2=0$ where the QCD calculation is reliable.

\begin{figure}[tb]
\hbox to\hsize{\hss\vbox{\offinterlineskip
\epsfxsize=0.8\hsize
\epsffile[60 210 550 550]{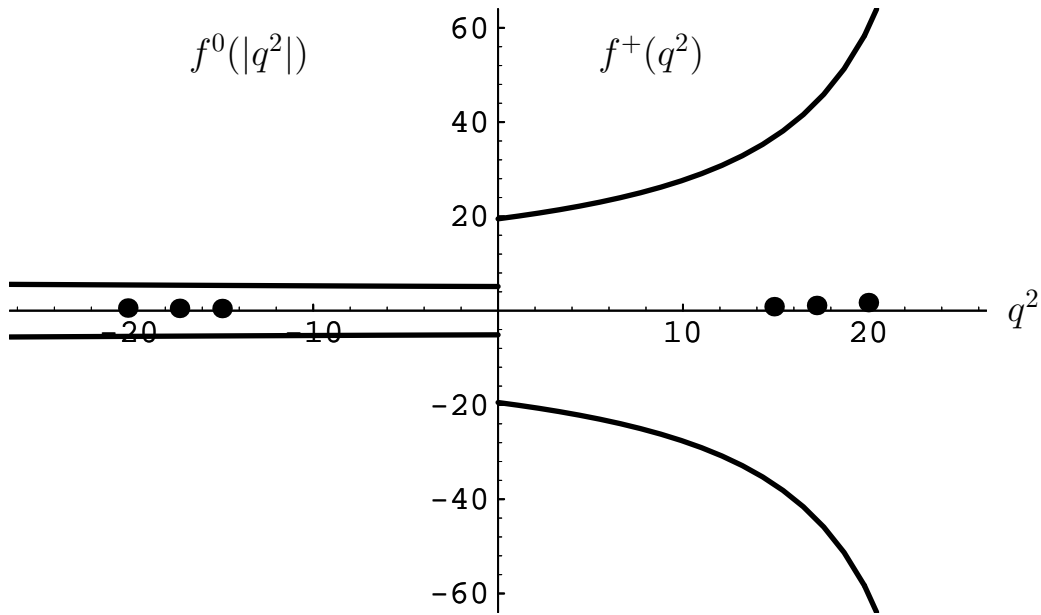}
\unit=0.8\hsize
\point 1.0 0.4 {\Large{$q^2$}}
\point 0.20 0.65 {\Large{$f^0(|q^2|)$}}
\point 0.60 0.65 {\Large{$f^+(q^2)$}}
}
\hss}
\caption[]{Bounds on 
$f^0(|q^2|)$ and $f^+(q^2)$ versus $q^2$ with $q^2$ in $\gev^2$.
$f^0(|q^2|)$ is plotted to the left of the vertical axis 
with $|q^2|$ increasing leftward from 0 to $q^2_{max}$ 
and $f^+(q^2)$ to the right
with $q^2$ increasing rightward from 0 to $q^2_{max}$. This way of
plotting the bounds is convenient, for it clearly shows whether
they satisfy the kinematical constraint of \protect\eq{kincst}.
The bounds plotted here do {\it not} implement this kinematical constraint 
nor do they take into account the
lattice results 
of the UKQCD Collaboration 
\protect\cite{jj} to which we have added systematic errors (points) (see
\protect\app{latpar}). Errors on these results are smaller than
the size of the points.
}
\label{nocstbnds}
\end{figure}

We plot the resulting, unconstrained bounds for $f^0(q^2)$ and $f^+(q^2)$
in \fig{nocstbnds}. 
These bounds are very loose and not interesting phenomenologically.
However, since the bounds on $f^0(q^2)$ are significantly better
than those on $f^+(q^2)$, it is clear that the
kinematical constraint of \eq{kincst} will significantly improve the
latter. 

\bigskip

\begin{figure}[tb]
\hbox to\hsize{\hss\vbox{\offinterlineskip
\epsfxsize=0.8\hsize
\epsffile[60 210 550 550]{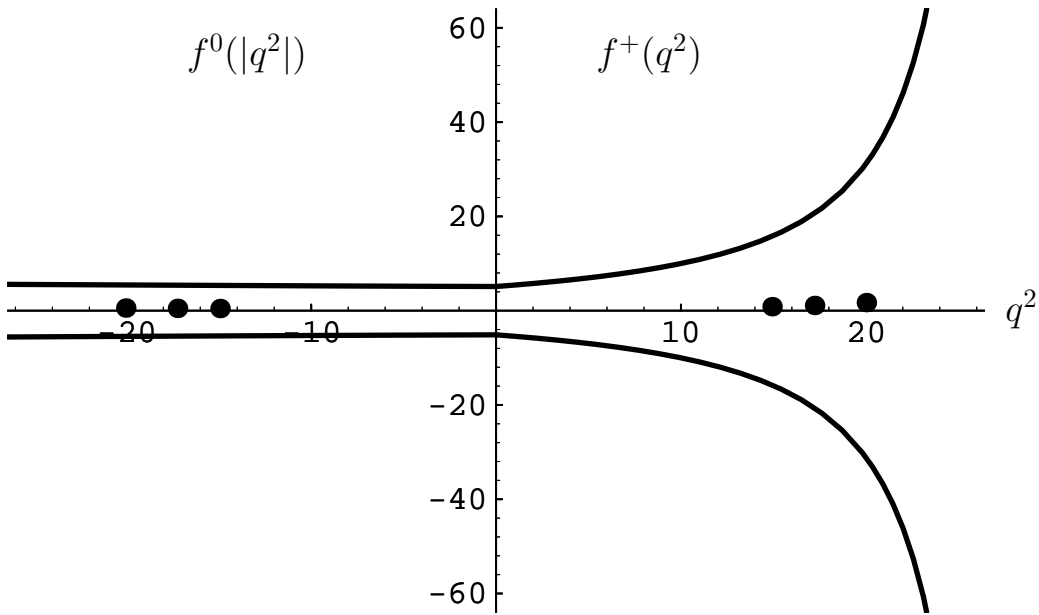}
\unit=0.8\hsize
\point 1.0 0.4 {\Large{$q^2$}}
\point 0.20 0.65 {\Large{$f^0(|q^2|)$}}
\point 0.60 0.65 {\Large{$f^+(q^2)$}}
}
\hss}
\caption[]{Bounds on 
$f^0(|q^2|)$ and $f^+(q^2)$ versus $q^2$ with $q^2$ in $\gev^2$,
as in \protect\fig{nocstbnds}.
These bounds implement the kinematical constraint of \protect\eq{kincst}
but do not take into account the
lattice results 
of the UKQCD Collaboration 
\protect\cite{jj} to which we have added systematic errors (points) (see
\protect\app{latpar}). Errors on these results are smaller than
the size of the points.
}
\label{kincstbnds}
\end{figure}

To implement this
kinematical constraint, we use the results of \sec{kin_cst}. For the
case at hand, $\phi_{lo}=F^0_{lo}(q^2{=}0|\vec 0)$ and
$\phi_{up}=F^0_{up}(q^2{=}0|\vec 0)$, where $F^0_{lo}(q^2{=}0|\vec 0)$
and $F^0_{up}(q^2{=}0|\vec 0)$ are the
bounds for $f^0(0)$ obtained without any additional input, i.e.
those plotted in \fig{nocstbnds}. The bounds on $f^0(q^2)$ are
unchanged while those on $f^+(q^2)$ are given by the versions of 
\eqs{upbndkin}{lobndkin} relevant for $f^+(q^2)$ with $\vec f^+=\vec 0$.
These new bounds are plotted in \fig{kincstbnds}. The kinematical constraint
is correctly implemented since the bounds agree at $q^2=0$. Furthermore,
the bound on $f^+(q^2)$
is significantly improved, especially at small and intermediate $q^2$
where phase space is largest. Thus, even though $f^0(q^2)$ does
not contribute to the total rate in the limit of vanishing lepton
mass, its bounds are important because, through the kinematical constraint,
they improve the bounds
on the form factor $f^+(q^2)$ which determines the rate.

\bigskip

We now consider the effect of using lattice results to further
constrain the bounds. The lattice calculation of \re{jj} provides the
values of $f^0(q^2)$ and $f^+(q^2)$ at three points above $q^2\simeq
15\gev^2$ (see \app{latpar}). Errors are taken into account as
described in \sec{errors} and below. Because of the large systematic
errors in the lattice results, it is unclear what form
the distribution ${\cal P}(\vec f^+,\vec f^0)$ should take. We have made
the rather conservative assumption: 1) that the results are gaussian
distributed about their central values with a variance given by the
large errors of \tab{latres}; 2) that these errors are uncorrelated. We
have checked that correlations have a tendency to reduce the range of
the bounds, as one would expect. We have also checked that a step function
distribution gives more constraining bounds.

\begin{figure}[tb]
\hbox to\hsize{\hss\vbox{\offinterlineskip
\epsfxsize=0.8\hsize
\epsffile[60 210 550 550]{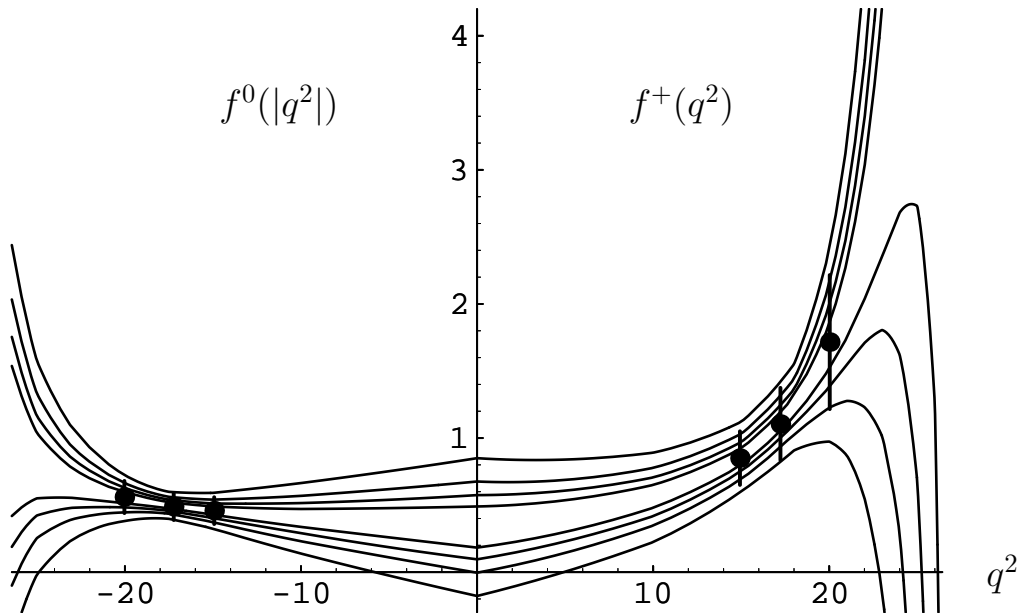}
\unit=0.8\hsize
\point 1.0 0.14 {\Large{$q^2$}}
\point 0.25 0.6 {\Large{$f^0(|q^2|)$}}
\point 0.65 0.6 {\Large{$f^+(q^2)$}}
}
\hss}
\caption[]{$f^0(|q^2|)$ and $f^+(q^2)$ versus $q^2$ with $q^2$ in $\gev^2$,
as in \protect\fig{nocstbnds}.  
The data points are the lattice results of the UKQCD Collaboration
\protect\cite{jj} to which we have added numerous systematic errors (see
\protect\app{latpar}). The pairs of curves are, from the outermost
to the innermost, our 90\%, 70\%, 50\% and 30\% model-independent QCD
bounds. These bounds implement the kinematical constraint of
\protect\eq{kincst} and take the
lattice results into account. As explained in the text, the fact that
they narrow less and less as the percentage is decreased is an
indication that the most probable behavior for the form factors is
within the 30\% bounds.}
\label{kinlatbnds}
\end{figure}

We use the probability of \eq{probp} for $f^+$ 
and the equivalent probability for
$f^0$ to define ``confidence level''
(CL) bounds for $f^+(q^2)$ and
$f^0(q^2)$. These $p\%$ bounds 
correspond to pairs of functions $(f_{lo}^+(q^2),f_{up}^+(q^2))$ and 
$(f_{lo}^0(q^2),f_{up}^0(q^2))$ such
that, for all $q^2$,
\be
{\cal P}_+(q^2;[f_{lo}^+(q^2),f_{up}^+(q^2)]|\vec f^+;\vec f^0)=0.01\,p
\ .\ee
To evaluate the various integrals involved, we perform a Monte Carlo
in which we generate 4000 independent bound samples for $f^+(q^2)$ and
$f^0(q^2)$. We find that the distributions for the various upper and
lower bounds at fixed $q^2$ are more or less bell shaped and nearly
symmetric. Since the width of a pair of bounds is typically quite
small compared to the width of these distributions, the latter are a
good guide as to the behavior of the probablility of \eq{probp} for
$f^+$ and the equivalent probability for $f^0$. Hence, to define
our $p\%$ bounds, we can consider the central $p\%$ of these
probabilities.  We find that the density in bounds increases as $p\%$
is varied from 95\% to about 30\%. This is confirmed by
\fig{kinlatbnds} where we plot bounds which have a $p\%$ ranging from
90\% for the outermost pair to 30\% for the innermost one in
decrements of 20\%: the space between neighboring bounds decreases 
as $p\%$ decreases. Beyond 30\%, it is unclear whether this
density continues to increase. In light of the
discussions of \sec{errors}, these $p\%$ bounds indicate that there
is at least a $p\%$ probability that $f^+(q^2)$ lie within
$[f_{lo}^+(q^2),f_{up}^+(q^2)]$ at each $q^2$ and similarly, that there is
at least a $p\%$ probability that $f^0(q^2)$ lie within
$[f_{lo}^0(q^2),f_{up}^0(q^2)]$. Thus, we conclude that the most
probable $q^2$-behavior of the form factors is within the 30\% bounds
though, of course, predictions outside these bounds but within the
70\% bounds are certainly not excluded.

We have checked that all of our results are stable with respect to the
number of Monte Carlo samples.  We have also constructed lattice-constrained
bounds for $f^+(q^2)$ and $f^0(q^2)$ that do not take into account the
kinematical constraint of \eq{kincst} and have found, not
surprisingly, that the resulting bounds on $f^+(q^2)$ are
significantly worse, especially for small $q^2$, while those for
$f^0(q^2)$ are not very different.

To get some idea of how strongly the bounds depend on the choice of
$Q^2$ , we have also determined the bounds for $Q^2=-16\gev^2$
assuming that the $\msbar$ results of \app{pert} for $\chi_L(Q^2)$ and
$\chi_T(Q^2)$ are reliable.  We find that the interval between the
70\% bounds on $f^0(q^2)$ shrinks by at most 3\% for $q^2\le 21\gev^2$
but shrinks by about 20\% around $q^2_{max}$.  Similarly, the bounds on
$f^+(q^2)$ shrink by at most 3\% for $q^2\le 20\gev^2$ but
shrink by about
25\% around $q^2_{max}$.

\section{Comparison of the Bounds with Various Parametrizations}
\label{comparison}

\begin{figure}[ht]
\vbox{
\vspace{5pt}
\offinterlineskip
\hbox to\hsize{\hfill
\epsfysize=0.37\hsize
\epsfxsize=0.60\hsize
\epsffile[75 200 550 550]{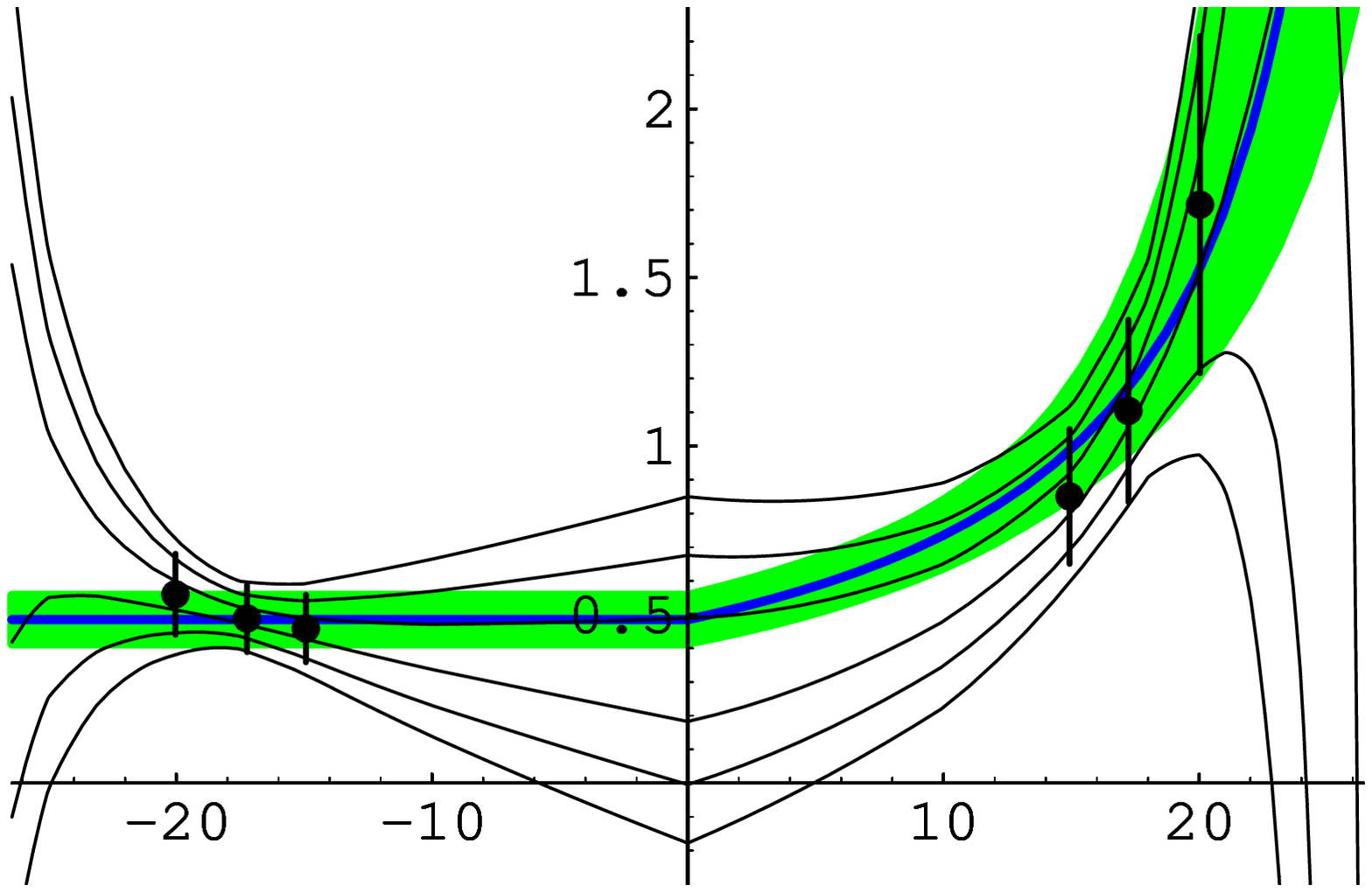}
\hfill
}%\kern1em
\hbox to\hsize{\hfill
\epsfysize=0.37\hsize
\epsfxsize=0.60\hsize
\epsffile[75 200 550 550]{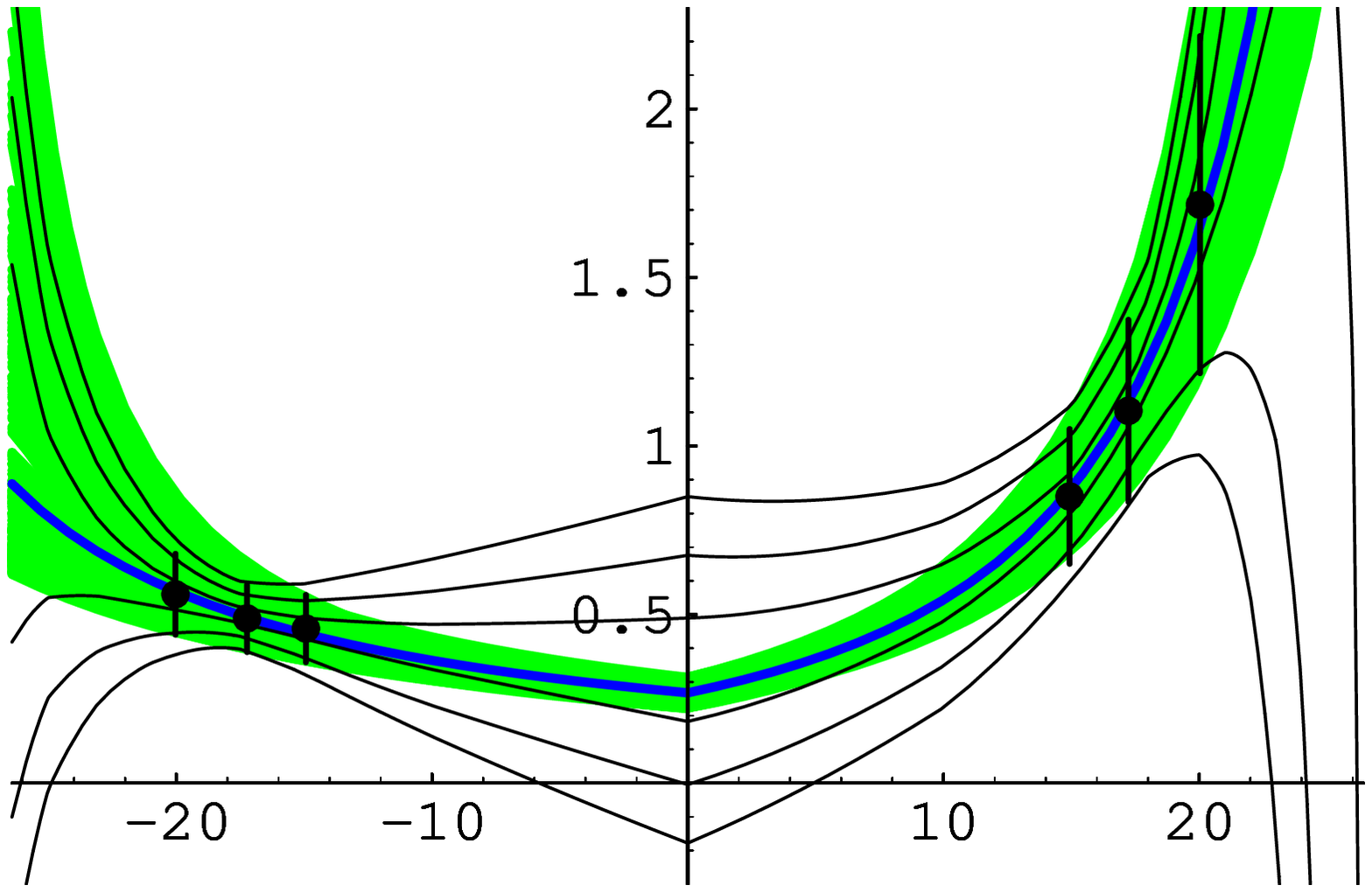}
\hfill
}%\kern1em
\hbox to\hsize{\hfill
\epsfysize=0.37\hsize
\epsfxsize=0.60\hsize
\epsffile[75 200 550 550]{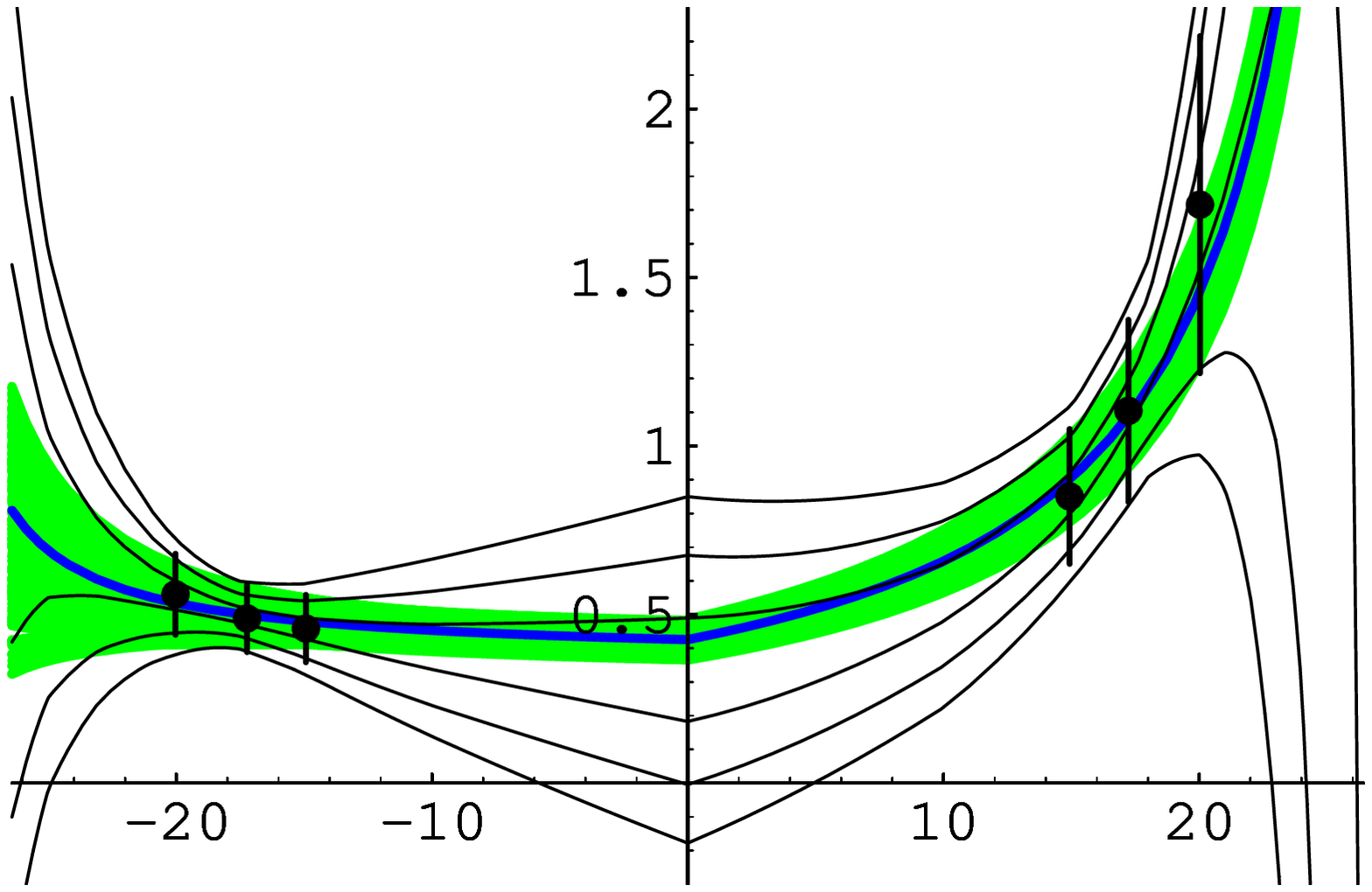}
\hfill}
\unit=0.4\hsize
\point 1.75 2.02 {\large{$q^2$}}
\point 1.75 1.07 {\large{$q^2$}}
\point 1.75 0.17 {\large{$q^2$}}

\point 0.77 2.65 {\large{$f^0(|q^2|)$}}
\point 0.77 1.75 {\large{$f^0(|q^2|)$}}
\point 0.77 0.80 {\large{$f^0(|q^2|)$}}

\point 1.42 2.65 {\large{$f^+(q^2)$}}
\point 1.42 1.75 {\large{$f^+(q^2)$}}
\point 1.42 0.80 {\large{$f^+(q^2)$}}

%\point 0.65 1.25 {\large{$(a)$}}
%\point 2.01 1.25 {\large{$(b)$}}
%\point 1.22 0.2 {\large{$(c)$}}
\point 1.2 1.99 {\large{$(a)$}}
\point 1.2 1.04 {\large{$(b)$}}
\point 1.2 0.12 {\large{$(c)$}}
}
\caption[]{$f^0(|q^2|)$ and $f^+(q^2)$ versus $q^2$ with $q^2$ in $\gev^2$,
as in \protect\fig{nocstbnds}.   
The data points are the lattice results of the UKQCD Collaboration
\protect\cite{jj} with additional systematic errors (see
\protect\app{latpar}). The pairs of fine solid curves are, from the outermost
to the innermost, our 90\%, 70\% and 30\% model-independent
QCD bounds. The thick curves are the results of fits and the bands, the
corresponding 68\% C.L. regions, for fits to
the three parametrizations described in \protect\sec{bkgnd}: $(a)$
``constant/pole''; $(b)$ ``pole/dipole''; $(c)$ ``fixed-pole'' with
the pole mass fixed to $m_{B^*}$.}
\label{figcomp}
\end{figure}

In \fig{figcomp} we plot the 90\%, 70\% and 30\% bounds
for $f^+(q^2)$ and $f^0(q^2)$ together with fits of the lattice data
to the ``constant/pole'', ``pole/dipole'' and ``fixed-pole''
parametrizations described in \sec{bkgnd}.  The results of
these fits are summarized in \tab{ffits}.  The bands which appear in the 
figure correspond to allowing the fit parameters to vary in their 
68\% CL region. Because of the limited
$q^2$-range of the lattice results for $f^0$, it is impossible to
reliably determine the two mass parameters $m_1$ and $m_2$ of the
``fixed-pole''parametrization of \eq{fzpole}.  Therefore, we have
allowed $m_1$ to vary and have taken $m_2$ to be the mass of the
nearest pole which can contribute to $f^0$, i.e.
$m_2=m_{B(0^+)}=5.46(30)\gev$~\cite{apebpi} where the $B(0^+)$ is the
first scalar excitation of the $B$ meson. The exact value of $m_2$ is
not, in fact, very important as the fit parameters of the
phenomenologically important form factor, $f^+(q^2)$, depend very
little on this value: we have varied $m_2$ in the range {}from $m_B$ to
$m_B+1\gev$ and have found that the fit to $f^+(q^2)$ changes by less
than one part in a thousand. What is important, however, is that with
both $m_1$ and $m_2$ on the order of $m_B$, the lattice data are fit
very nicely and the resulting curves lie within the QCD bounds.
\begin{table}[tb]
\begin{center}
\begin{tabular}{|c||c|c|c|c|c|} \hline
fit type & $f^+(0)$ & ``$m_{B^*}$'' or 
$m_+\,(\gev)$ & $m_1$ or $m_o\,(\gev)$ & $\chi^2/d.o.f.$\\ \hline\hline
fixed-pole & $0.43\pm 0.04$ 
& $m_{B^*}$ & $5.9\pm 0.3$ & 0.4/4\\ 
 & $0.39\pm 0.04$ 
& 5.17 & $6.2\pm 0.4$ & 0.2/4\\ 
 & $0.52\pm 0.05$ 
& 5.9 & $5.4\pm 0.2$ & 1.7/4\\ \hline
cst/pole & $0.48\pm 0.05$ & $5.42\pm 0.26$ & & 1.0/4\\ \hline
pole/dipole & $0.27\pm 0.03$ & $5.79\pm 0.19$ & $6.2\pm 0.4$ 
& 0.1/3 \\ \hline
\end{tabular}
\caption{Fits of the lattice results for $f^+(q^2)$ and $f^0(q^2)$ 
to the various parametrizations described in \protect\sec{bkgnd}:
fixed-pole (\protect\eqs{fppole}{fzpole}); cst/pole
(\protect\eq{npole} with $n=0$); pole/dipole (\protect\eq{npole}
with $n=1$).  All of these parametrizations are consistent with the
kinematical constraint of \protect\eq{kincst} (i.e. $f^+(0)=f^0(0)$)
and heavy-quark scaling laws.
\label{ffits}}
\end{center}
\end{table}

As evidenced by the low values of $\chi^2/dof$ obtained for all three
sets of fits, the lattice results alone cannot effectively
discriminate amongst the different parametrizations, at least when
systematic errors are taken into account before performing these fits.
As mentioned in the Introduction, this early inclusion of systematic
errors, some of which are independent of $q^2$, may be partially
responsible for this lack of effectiveness. However,
since the relative proportions of $q^2$-independent and
$q^2$-dependent systematic errors are not known, we have chosen to be
conservative by including all errors {}from the very beginning.  

Our lattice constrained bounds do not exclude unambiguously any of the
parametrizations either: the best fits of all three sets lie within
our 70\% bounds. However, substantial fractions of the parameter
values allowed at the 68\% CL by the fits are excluded by our 70\% and
even 90\% bounds. 
\footnote{Some of the pole mass values allowed in the 68\% CL 
ellipse for the ``constant/pole''
fit lead are smaller than $q^2_{max}$ 
and therefore lead to an $f^+(q^2)$ which
diverges.}
Furthermore the best ``constant/pole'' fit
lies outside the 30\% bounds for a large range of $q^2$. 
In fact, only the best ``pole/dipole'' fit lies within these 30\%
bounds for all
$q^2$. Since these
bounds delimit the region of most probable values for the form
factors, the ``constant/pole'' parametrization is least likely to be an
accurate description of the form factors while the ``pole/dipole'' form
is the most probable parametrization. This is confirmed
by the 
the results of \sec{brbnds} where we find that only the best ``pole/dipole''
fit yields a value of the total rate that is consistent with
our 30\% bounds on that rate: the best ``fixed-pole'' fit rate only lies
within the 50\% bounds and the best ``constant/pole'' fit rate, 
only within the 70\% bounds.
However, the bounds will have to improve before
a firm conclusion as to the preferred $q^2$-behavior of the form
factors can be drawn.

For completeness, we have also performed a fit of the lattice results
to the ``fixed-pole'' parametrization with different values of
the pole mass (i.e.
$m_{B^*}$ in \eq{fppole}). We find that the pole
masses allowed range {}from $5.17\gev$ to $5.9\gev$.  Outside this
range, the pole prediction lies either above
(``$m_{B^*}$''$\le 5.16\gev$) or below (``$m_{B^*}$''$\ge 6.0\gev$) our
70\% bounds on $f^+(q^2)$. We have further allowed the pole mass to vary
freely. 
We find $f^+(0)=0.35(3)$,
$m_1=6.6(6)\gev$ and ``$m_{B^*}$''$=5.01(14)\gev$ with a $\chi^2/dof=
0.004/3$. As this fit indicates, the lattice data alone may favor a
pole mass and a value of $f^+(0)$ slightly smaller than those obtained
in the ``fixed-pole'' fits of \tab{ffits}.  Thus, $f^+(q^2)$ could be
dominated by the $B^*$ pole for $q^2$ greater than $15-20\gev^2$
but have a $q^2$-dependence determined by a smaller pole mass for
smaller $q^2$, as suggested in \re{VlBB95}.  Such a behavior is
perfectly compatible with our bounds as are, of course, many
other more complicated behaviors. 

We delay making comparisons of our results for $f^+$ with those
of other authors until \sec{brbnds} where we also compare 
predictions for the total rate. In particular, we collect 
the predictions of other authors for $f^+(0)$ in \tab{ratecomp}.

\clearpage
%\pagebreak

%

\clearpage

\section{The Couplings $g_{B^*B\pi}$ and $g$}
\label{cplngbnds}

Having obtained bounds on $f^+(q^2)$, we can use
\eqs{resbndone}{resbndtwo} to constrain the $B^*B\pi$ coupling,
$g_{B^*B\pi}$. Using \eq{fp} and $f_{B^*}=33\er{4}{4}$ (see comments
after \eq{jqsqp}), we find
\be
-8.9\pm 1.1\le -6.2\pm 0.8\le -3.4\pm 0.4\le g\le 6.7\pm 0.8
\le 9.3\pm 1.1\le 11.6\pm 1.4
\label{gbnd}
\ee
and
\be
-724\pm 88\le -505\pm 61\le -275\pm 33
\le g_{B^{*+}B^0\pi^+} \le 544\pm 66\le 751\pm 91\le 942\pm 114
\ ,
\label{gbsbpibnd}
\ee
where the outer limits correspond to a CL of 95\%, the next interval
inward, to a CL of 70\% and the smallest interval to a CL of 30\%. The
errors come from the uncertainties on $f_{B^*}$.  These bounds are weak
because we do not have lattice results for $f^+(q^2)$ very near
$q^2_{max}$. This will be remedied, though, as soon as the lattice
calculation of \re{jj} is repeated with a sufficient number of light
quark masses to enable a numerical chiral extrapolation of the form
factors (see \app{latpar} for details) and thus provide determinations
of the form factors closer to $q^2_{max}$. 

For completeness, we also give the results for these couplings
obtained under the assumption of pure $B^*$ dominance.  As we have
seen, $B^*$ dominance describes the lattice results well and is
consistent with our bounds. We find:
\be
g=0.35\pm 0.05
\label{gpole}
\ee
and
\be
g_{B^{*+}B^0\pi^+}=28\pm 4
\ .
\label{gbbspipole}
\ee
Both these results and the bounds include statistical and systematic
errors since the lattice results used to obtain these results do (see
\app{latpar} for details). However, it is important to note
that only the bounds of \eqs{gbnd}{gbsbpibnd} 
are model-independent and that the errors 
quoted in \eqs{gpole}{gbbspipole} do not include the possible
deviation of the form factor $f^+(q^2)$ from pure $B^*$-pole
behavior. The results of \eqs{gpole}{gbbspipole} must therefore
be treated cautiously.

Nevertheless, the pole result for $g_{B^{*+}B^0\pi^+}$ compares very
favorably with the light-cone sumrule prediction of Belyaev {\it et
al.}~\cite{VlBB95} who find $g_{B^{*+}B^0\pi^+}=29\pm3$. It is rather
high, however, compared to the soft-pion sumrule result of Colangelo
{\it et al.}~\cite{PiCD94}, $g_{B^{*+}B^0\pi^+}=15\pm4$, and to the
QCD double-moment spectral sumrule result of Dosch {\it et al.},
$g_{B^{*+}B^0\pi^+}=14\pm4$~\cite{HaDN95}.  As pointed out in
\re{VlBB95}, the sumrule of \re{PiCD94} can be derived {}from the
light-cone sumrule of \re{VlBB95}. Though the authors of \re{VlBB95}
and \re{PiCD94} agree on the sumrule, they disagree on its evaluation,
the former finding $g_{B^{*+}B^0\pi^+}=28\pm6$ in the soft-pion limit.
For a discussion of this disagreement as well as a good summary of
recent determinations of $g_{B^*B\pi}$, please see
\re{VlBB95}.

\section{Bounds on the Rate, Comparisons with other Predictions and $|V_{ub}|$}
\label{brbnds}

Our bounds on $f^+(q^2)$ enable us to constrain the rate
$d\Gamma\l(\bpi\r)/dq^2$. In the limit of vanishing lepton
mass, which is an excellent approximation for $\ell=e,\,\mu$, we have
\be
\frac{d\Gamma}{dq^2}(\bpi)=\frac{G_F^2|V_{ub}|^2}{192\pi^3 m_B^3}
\,\lambda^{3/2}(q^2)\,|f^+(q^2)|^2
\ ,
\label{ddrate}
\ee
where $\lambda(q^2)=(m_B^2+m_\pi^2-q^2)^2-4m_B^2m_\pi^2$ and $0\le q^2\le
q^2_{max}$. 

As was done for $f^+(q^2)$ and $f^0(q^2)$ in \sec{errors}, we
define the minimum probability that the total rate for
$\bpi$ decays takes a value inside an interval
$[\Gamma_{lo},\Gamma_{up}]$. This probability is given by
\bea
{\cal P}_\Gamma([\Gamma_{lo},\Gamma_{up}]|\vec f^+;\vec f^0) 
\equiv \frac{1}{Z(\vec f^+;\vec f^0) }
\int & d\mu(\vec f^+;\vec f^0)\,\,{\cal P}_{in}(\vec f^+,\vec f^0)\,\,
\theta\l(\Gamma_{up}-G_{up}\l(\vec f^+,f^+(0)\in[\phi_{lo},\phi_{up}]\r)\r)
\nonumber\\
&\times\,
\theta\l(G_{lo}\l(\vec f^+,f^+(0)\in[\phi_{lo},\phi_{up}]\r)-\Gamma_{lo}\r)
\label{probgamma}
\ ,\eea
where
$$
G_{up}(\vec f^+,f^+(0)\in[\phi_{lo},\phi_{up}])
\equiv \frac{G_F^2|V_{ub}|^2}{192\pi^3 m_B^3}\int_0^{q^2_{max}} dq^2\,
\lambda^{3/2}(q^2)
$$
\be
\times\quad
\mbox{Max}\l\{\l|F_{up}^+(q^2|\vec f^+,f^+(0)\in[\phi_{lo},\phi_{up}])\r|^2
\, ,\,
\l|F_{lo}^+(q^2|\vec f^+,f^+(0)\in[\phi_{lo},\phi_{up}])\r|^2
\r\}
\label{ratebndup}
\ee
and
$$
G_{lo}(\vec f^+,f^+(0)\in[\phi_{lo},\phi_{up}])
\equiv \frac{G_F^2|V_{ub}|^2}{192\pi^3 m_B^3}\int_0^{q^2_{max}} dq^2\,
\lambda^{3/2}(q^2)
$$
\be
\times\quad
\l\{
\begin{array}{l}
0\, ,\quad  \mbox{if $F_{lo}^+(q^2|\vec f^+,f^+(0)\in[\phi_{lo},\phi_{up}])
\le 0 \le F_{up}^+(q^2|\vec f^+,f^+(0)\in[\phi_{lo},\phi_{up}])$}\\
\mbox{Min}\l\{\l|F_{up}^+(q^2|\vec f^+,f^+(0)\in[\phi_{lo},\phi_{up}])\r|^2
\, ,\,
\l|F_{lo}^+(q^2|\vec f^+,f^+(0)\in[\phi_{lo},\phi_{up}])\r|^2
\r\},\quad  \mbox{otherwise,}
\end{array}
\r.
\label{ratebndlo}
\ee
where $F_{lo}^+(\cdots)$ and $F_{up}^+(\cdots)$ are the bounds on
$f^+(q^2)$ defined in \sec{kin_cst}. The above expressions for the bounds
on the rate are involved because the bounds on $f^+(q^2)$ can change
signs with $q^2$.

Now, using the probability of \eq{probgamma}, we can define $p\%$ CL
bounds for the rate. To calculate this probability, we use the results
of the Monte Carlo for the bounds on the form factors of \sec{bounds}.
Since, again, the width of a pair of bounds is typically quite small
compared to the width of the distributions for upper and lower bounds,
the latter are a good guide as to the behavior of the probablility of
\eq{probgamma}.  We find that these distributions are bell shaped, but
definitely peaked toward lower values of the rate. Therefore, instead
of considering the central $p\%$ for our defining our bounds, we
resort to a different procedure. Starting at the egdes of our
collection of bounds, we work our way inward toward the most probable
pairs of bounds by taking steps of constant size.  Because the density
of bounds increases much faster as we approach the most probable pairs
of bounds from below than from above, we take the upward step to be
much smaller than the downward step so that the increase in bound
density is more or less equal in either direction. We use
$0.2|V_{ub}|^2\, ps^{-1}$ upward steps and $1.0|V_{ub}|^2\, ps^{-1}$
downward steps, because they are a good compomise between size and 
upward and downward balance. We then call $p\%$ bound
the smallest interval obtained by taking these steps which includes at
least $p\%$ of the bounds. Our results depend little on the
exact size of the steps or even on the procedure. For instance, taking
constant percentage steps of 2\% or more on whichever side of the
collection of bounds is less dense gives very similar
results.  In all cases, we find that this density increases until
about only 30\% of the bounds are left. Even though the density does
continue to increase slightly beyond that point, we prefer to limit
ourselves to the statement that the most probable values for the rate
is somewhere within these 30\% bounds.  The results given by our
procedure are well
corroborated by the distributions of lower and upper bounds mentioned
above.

We present these results in \tab{ratecomp}. For comparison, we also
collect the predictions of various authors for the total rate. We
specify the $q^2$-range for which their calculations are valid and the
assumptions made to extend the results to all of phase space. We further
provide our and their predictions for $f^+(0)$. The spread of results
for the rate is very large, the highest and lowest predictions
differing by a factor of about 7. Such a spread implies a factor of
about 2.6
uncertainty in the determination of $|V_{ub}|$ and clearly shows the
necessity for a completely model-independent prediction.
\clearpage
\begin{table}[tb]
\begin{center}
\begin{tabular}{|c|c|c|c|} \hline
Reference & $\Gamma\l(\bpi\r)$ & $f^+(0)$ & Details\\ \hline
This work & $2.4\to 28$ & $-0.26\to 0.92$ & 95\% CL\\
& $2.8\to 24$ & $-0.18\to 0.85$ & 90\% CL\\
& $3.6\to 17$ & $0.00\to 0.68$ & 70\% CL\\
& $4.4\to 13$ & $0.10\to 0.57$ & 50\% CL\\
& $4.8\to 10$ & $0.18\to 0.49$ & 30\% CL\\
\hline
QM~\cite[(WSB)]{WSB85} & $7.4\pm 1.6$ & $0.33\pm 0.06$ & 
$q^2{=}0$; $B^*$ pole
\\ \hline
QM~\cite[(ISGW)]{NaIS89} & 2.1 & 0.09 & $\sim q^2_{max}$; 
exponential\\ \hline
QM~\cite[(ISGW2)]{DaSI95} & 9.6 & & $\sim q^2_{max}$; dipole\\ \hline
QM~\cite{PaOX95} & & $0.26$ & $q^2{=}0$ \\ \hline
QM~\cite{ChCHZ96} & & $0.24-0.29$ & QM+$B^*$ pole 
\\ \hline
QM~\cite{IGNS96} & 9.6--15.2 & $0.29-0.46$ & $q^2{=}0\sim 20\gev^2$ \\ \hline
SR2~\cite
{CeDP88} & $14.5\pm 5.9$ & $0.4\pm 0.1$ & $q^2_{max}$; 
modified $B^*$ pole\\ \hline
SR$2+3$~\cite{AO89} & 4.5--9.0 & $0.27\pm 0.05$ & $q^2{=}0$ and $q^2_{max}$; \\
& & & modified $B^*$ pole\\ \hline
SR3~\cite{StN95} & $3.60\pm 0.65$ & $0.23\pm 0.02$ & 
$q^2=0\sim 5\gev^2$\\
& & & $+$ $B^*$ pole for rate\\ 
\hline
SR3~\cite{PaB93} & $5.1\pm 1.1$ & $0.26\pm 0.02$ & 
$q^2{=}0\sim 16-20\gev^2$\\
& & & $+$ pole fit w/\\
& & & $m_{pole}{=}5.25(10)\gev$\\ \hline
LCSR~\cite{VlBK93,VlBB95,AlKR95} & 
$8.1$  & $0.24-0.29$ & $q^2{=}0\sim 15\gev^2$, then $B^*$ pole \\ \hline
LAT~\cite[(UKQCD)]{jj}$^\dagger$
& $7\pm 1$
 & 0.21 -- 0.27 & $q^2\simeq 15\sim 20\gev^2$\\
& & & $+$ ``pole/dipole'' fit\\ \hline
LAT~\cite[(ELC)]{ELC94} & $9\pm 6$ & 0.10--0.49 & single point at 
$q^2\simeq 18\gev^2$\\
& & & $+$ pole w/ $m_{pole}{=}5.29(1)\gev$\\ \hline
LAT~\cite[(APE)]{apebpi} & $8\pm 4$ & 0.23--0.43 & single point at 
$q^2\simeq 20.4\gev^2$\\
& & & $+$ pole w/ $m_{pole}{=}5.32(1)\gev$\\ \hline
\end{tabular}
\caption{
Comparison of different theoretical predictions for $f^+(0)$ and 
the total 
rate for $\bpi$ decays (in units of $|V_{ub}|^2\,ps^{-1}$).
The results in the first row block correspond to our 95\%, 90\%, 70\%,
50\% and 30\% 
bounds on these quantities as described in the present
section and \protect\sec{bounds}.  In column 1, QM stands for quark
model; SR2 for two-point function QCD spectral sumrules; SR3 for
three-point function QCD spectral sumrules; LCSR for light-cone
sumrules; and LAT stands for lattice. 
The last column specifies the
range of $q^2$ for which the calculations are reliable and the
assumptions and/or fits made to extend the results to the full
kinematical range.  
\label{ratecomp}}
\begin{minipage}[t]{\textwidth}
\protect\rule{7cm}{0.1mm}
\protect\\
{\small $^\dagger$ The value for the rate 
differs from the one quoted in \protect\re{jj}.
Here, the phase space integral is performed from $q^2=0$ to 
$q^2_{max}=26.4\gev^2$ assuming flavor symmetry in the 
light, active and spectator quarks 
as discussed in \protect\app{latpar}. Neither the rate nor
the value of $f^+(0)$ include systematic errors\protect\cite{drb}.}
\end{minipage}
\end{center}
\end{table}

\clearpage
%\medskip

Our bounds on $f^+(0)$ are not very strong and at the 70\% CL they
are compatible with all other predictions. As discussed in \sec{bounds}, 
the most probable value for $f^+(0)$ lies within the 30\% CL bounds,
i.e. in the interval
$[0.20,0.48]$, which is still consistent
with most predictions for $f^+(0)$, except the ISGW result of \re{NaIS89}.

Our bounds on the rate, on the other hand, do make some of the other
predictions quite unlikely. The ISGW model 
result of \re{NaIS89} is excluded at
the 95\% CL; the central value of the sumrule result of 
\re{StN92}, at the 70\% CL, and the
value at the tip of the error bar, at the 50\% CL. All other results
are compatible with our bounds at the 70\% CL, 
including the two lattice results of
\re{ELC94} and
\re{apebpi}. They
were obtained {}from a single measurement of $f^+$ for a $q^2\sim
18-20\gev^2$, assuming $B^*$ dominance with $m_{B^*}$ taken {}from a
lattice computation with the same gluon configurations. Though the
errors are mainly statistical or due to a heavy-quark mass
extrapolation of the form factors similar to the one described in
\app{latpar}, they were made as large as possible
to try to accomodate the fact that the $q^2$-dependence of the form factor is
not determined by the calculation. 

Also, for $q^2\ne 0$ our bounds on $f^+(q^2)$ have more to say, some of which
could be inferred, in a different language, 
from a $\chi^2$ analysis with the lattice
data alone. Considering only central values for the predictions
of other authors, we find,
for instance, that the sumrule result \re{PaB93} is excluded by our 70\%
lower bound for over one-third of the kinematical range from 
$q^2\simeq 12\gev^2$ to $q^2\simeq 22\gev^2$ and by the 90\%
lower bound from $q^2\simeq 15\gev^2$ to $q^2\simeq 20\gev^2$. 
The WSB model, or equivalently the
lattice result of \re{apebpi}
\footnote{It is important to note that the actual lattice number
from which this result is obtained is compatible, within errors, with those of
\protect\app{latpar}.}, 
is only excluded by the 70\% lower bound
for less than a fifth of the kinematical range, between $q^2\simeq
15\gev^2$ and $q^2\simeq 21\gev^2$.  The light-cone sumrule result of
\re{VlBB95} agrees with our bounds extremely well since it very nearly
lies within our 30\% bounds for $f^+(q^2)$ 
for all $q^2$. Very good agreement is also
found for the light-front quark model, non-relativistic hamiltonian
(NR) and harmonic oscillator wavefunction (HO) results of \re{IGNS96}
which lie within our 50\% bounds for all $q^2$.  Furthermore, 
the Godfrey-Isgur
(GI) result of \re{IGNS96} agrees well with our bounds since
it very nearly lies within the 70\% bounds for all $q^2$.  However, 
this agreement
with the results of \re{IGNS96}
implies that the $B^*\pi$ contribution to the $B$-meson
wavefunction, which is neglected in \re{IGNS96}, can 
only become significant
beyond $q^2\sim 20\gev^2$ where our bounds on $f^+(q^2)$
allow much more singular
behavior than that given by the results of \re{IGNS96}. 
This observation is corroborated by the 
light-front quark model calculations of
\re{ChCHZ96}, since our bounds on $f^+(q^2)$ 
favor the result which relies on the smallest value of the coupling
$g$ that these authors consider.

We have also computed the rates obtained from the fits to the lattice
data of various parametrizations for the form factors performed in
\sec{comparison}. The results are summarized in \tab{fitrates}. As these
results show, only the ``pole/dipole'' parametrization gives a central
value for the rate which lies within our 30\% bounds which, together
with the observations of \sec{comparison}, indicates
that, of the forms we have tried, it appears to be the most probable
one. Then comes the ``fixed-pole'' prediction which lies within the
50\% bounds and finally the ``constant-pole'' rate which is excluded
by the 50\% bounds but lies within the 70\% bounds. However, once again,
the bounds will have to improve 
before a firm conclusion as to the preferred
$q^2$-behavior of the form factors can be drawn.
\begin{table}[tb]
\begin{center}
\begin{tabular}{|c||c|} \hline
fit type & $\Gamma\l(\bpi\r)$ in units of $|V_{ub}|^2\,ps^{-1}$\\ \hline\hline
cst/pole & $15.2+7.5-3.1$ \\ \hline
pole/dipole & $8.5+3.0-1.9$ \\ \hline
fixed-pole & $12.3+2.5-2.2$\\ \hline
\end{tabular}
\caption{Values for the rate obtained from the fits described in
\protect\sec{comparison}. The error bars are 70\% errors.
\label{fitrates}}
\end{center}
\end{table}

Finally, the most probable
value for the rate is somewhere in the interval 
$[4.8,10]|V_{ub}|^2\, ps^{-1}$ which
corresponds to the 30\% CL bounds.
Thus, we can summarize our results for the rate as
\be
\Gamma\l(\bpi\r) = (4.8\leftrightarrow 10)\errr{7.0}{1.2} 
|V_{ub}|^2\, ps^{-1}
\label{rateres}
\ee
where the errors are obtained from the 70\% CL bounds.
\footnote{These errors are not gaussian.}
Because
we have been generous in our estimate of possible 
systematic errors, 
the theoretical error in \eq{rateres} is probably an overestimate
and the 60\% CL bounds or even perhaps the 50\% CL bounds may give 
a better estimate of this theoretical error. For instance, the 
estimate given by the 50\% CL bounds is:
\be
\Gamma\l(\bpi\r) = (4.8\leftrightarrow 10)\errr{3.0}{0.4} 
|V_{ub}|^2\, ps^{-1}
\ .\label{rateresless}
\ee

\bigskip

Now, the CLEO Collaboration very recently measured the
branching ratio for $\bpi$ decays~\cite{bpicleo}. They found
\be
\br{\bpi} = \l\{\begin{array}{ll}
(1.63\pm 0.46\pm 0.34)\times 10^{-4} & \mbox{WSB}\\
(1.34\pm 0.35\pm 0.28)\times 10^{-4} & \mbox{ISGW}
\end{array}\r.\ ,
\ee
where experimental efficiencies are determined
with the WSB model~\cite{WSB85} and 
the ISGW model~\cite{NaIS89}, respectively. Because of the model
dependence of their results, a precise statement about $|V_{ub}|$ is not
yet possible. However, to illustrate the sort of accuracy our 
results lead to, we proceed to extract $|V_{ub}|$ {}from these measurements. 
Because our bounds favor
very strongly the WSB model predictions for $f^+$ and the rate over
those of the ISGW model, we consider only the WSB measurement.
Using the result of \eq{rateres}, we find
\be
|V_{ub}|\,\sqrt{\tau_{\bar B^0}/1.56\,ps} = 
(0.0032\leftrightarrow 0.0047)+0.0007-0.0007\pm 0.0007 \quad \mbox{(WSB)}
\ ,
\label{vubres}
\ee
where the first set of errors is theoretical (non-gaussian)
and the second experimental
(statistical and systematic combined in quadrature on the average value
of $|V_{ub}|$ given by the 30\% CL results). The value of $\tau_{\bar B^0}$
we have taken is $1.56(5)\,ps$~\cite{IKr96}.
Using the less conservative result of 
\eq{rateresless} for the total rate, we find
\be
|V_{ub}|\,\sqrt{\tau_{\bar B^0}/1.56\,ps} = 
(0.0032\leftrightarrow 0.0047)+0.0002-0.0004\pm 0.0007 \quad \mbox{(WSB)}
\ ,
\label{vubresless}
\ee
where the origin of the errors is the same as in \eq{vubres}.
Finally, using the 90\% CL bounds on the rate, we get
\be
0.0021\pm 0.0004
\le |V_{ub}|\,\sqrt{\tau_{\bar B^0}/1.56\,ps} \le
0.0061\pm 0.0011 \quad \mbox{(WSB)}
\ ,
\label{vubbnd90}
\ee
and using the the 95\% CL bounds,
\be
0.0019\pm 0.0003
\le |V_{ub}|\,\sqrt{\tau_{\bar B^0}/1.56\,ps} \le
0.0066\pm 0.0012 \quad \mbox{(WSB)}
\ ,
\label{vubbnd95}
\ee
where the errors shown on the bounds are the experimental errors.

The determination of $|V_{ub}|$ given in \eq{vubres} has a theoretical
error of 37\%, computed by taking the difference of the 70\% CL
bound results over their sum.
\footnote{Though this error is not gaussian, it gives a good
idea of the sort of accuracy achieved.} 
Though this error is by no means negligible, it
is nevertheless quite reasonable especially since the result of
\eq{vubres} is
completely model-independent and is obtained from lattice data
which include a conservative range of systematic errors (please
see \app{latpar} for details). The determination of $|V_{ub}|$ 
given in \eq{vubresless}, obtained under less conservative assumptions
(i.e. from the 50\% CL bounds),
has a theoretical error of 27\%.

\section{Conclusion}
\label{ccl}

We have combined lattice results, a kinematical constraint and QCD
dispersion relations to derive model-independent bounds on the form
factors relevant for semileptonic $\bpi$ decays. Even though the 
contribution of the form
factor $f^0$ to the rate is negligible, we have found
its bounds useful for constraining $f^0(0)=f^+(0)$, thereby making the
bounds on $f^+(q^2)$ stronger. And although our bounds 
do not unambiguously exclude any of the parametrizations for the
$q^2$-dependence of the form factors that we have tried,
\footnote{These parametrizations fit the lattice results
well and are consistent with heavy-quark scaling laws and the kinematical
constraint.} 
they do limit the range of possible parameter values
and indicate that some of the parametrizations are more probable than others.
We find, for instance, that the ``pole-dipole'' parametrization of
\sec{bkgnd} appears to be more likely than the ``fixed-pole'' parametrization 
which, in turn, appears more probable that the ``constant-pole''
parametrization. We also find excellent agreement with the light-cone sumrule
results of \re{VlBB95} for $f^+(q^2)$ which almost lie within our 30\%
bounds for all $q^2$. 
Agreement with our bounds for $f^+(q^2)$ is also very good for
the non-relativistic hamiltonian (NR) and 
the harmonic oscillator wavefunction (HO) 
results of the light-front quark model 
calculations of \re{IGNS96} which both lie within our 50\% bounds for all
$q^2$. Further good agreement is found with the Godfrey-Isgur
(GI) result of \re{IGNS96} which very 
nearly lies within our 70\% bounds for all $q^2$. This agreement
with the results of \re{IGNS96}
implies that the $B^*\pi$ contribution to the $B$-meson
wavefunction, which is neglected in \re{IGNS96}, can 
only become significant
beyond $q^2\sim 20\gev^2$ where our bounds on $f^+(q^2)$ 
allow much more singular
behavior than that given by the results of \re{IGNS96}. 
Agreement of 
other predictions with our bounds ranges from slightly less 
good to much less good. However, the bounds will have to improve before a firm
conclusion as to the preferred $q^2$-behavior of the form factors can
be drawn. 

We also use our bounds, as well as the vector dominance results, to
determine the $B^*B\pi$ coupling and the coupling $g$ which appears in
Heavy-Meson Chiral lagrangians at leading order. While our bounds on
these couplings are quite weak, our ``fixed-pole'' fit to the form
factors yields $g=0.35\pm 0.05$ (\eq{gpole}) and
$g_{B^{*+}B^0\pi^+}=28\pm 4$ (\eq{gbbspipole}).  It is important to
note, however, that the ``fixed-pole'' results rely on the assumption
that $f^+$ is dominated by the $B^*$ pole over the full kinematical
domain.

We further derive bounds on the total rate.  These bounds
disfavor some quark model and sumrule predictions with some
certainty. At the 70\%
CL, they also enable the extraction of $|V_{ub}|$ {}from the
corresponding experimental results with a theoretical error of 37\%.
Since we have been generous in our estimate of possible systematic
errors, this 37\% is probably an overestimate. It can be reduced
to 27\% if one is willing to accept the 50\% CL bounds 
as a good estimate
of the theoretical uncertainties. More 
generally, this extraction is complementary to the lattice
determination of $|V_{ub}|$ from the differential decay rate for
$\brho$ decays suggested in \cite{JoFN9506}.  Furthermore, the
preliminary determination of the branching ratio for $\bpi$ decays by
the CLEO collaboration~\cite{bpicleo} suffers from a rather strong
model dependence.  Perhaps this model dependence could be reduced by
using our results.

As mentioned in the Introduction and described in \app{latpar}, we
assume flavor symmetry in the light, active and spectator quarks in
using the UKQCD Collaboration lattice results of \re{jj} for
$f^+(q^2)$ and $f^0(q^2)$.  Although there is evidence
that such a symmetry is reliable
for the values of $q^2$ we consider,
we add large systematic errors to the data to cover 
possible violations. We further add a very large range of 
errors to account for other
possible systematic effects.
With these systematic errors (and in some cases even without)
the lattice results of \protect\tab{latres} are compatible
with the very few 
chirally extrapolated points of Refs. \protect\cite{ELC94} and
\protect\cite{apebpi} within one standard deviation or less.
Of course, our whole
analysis should be repeated once a complete set of reliable chirally
extrapolated results are available.

We add all of these systematic errors to the lattice results before
performing our analysis.  We believe that such a procedure is sensible
in phenomenological applications when systematic uncertainties
dominate over statistical ones.  As discussed in the Introduction and
in \sec{comparison} this gives a more conservative representation of the
accuracy of the lattice results. We are thus confident that our final
results are correct within these large errors. 

We have also investigated the effect of reducing lattice errors.
With errors reduced by a factor 4, the corresponding lattice
constrained bounds would certainly discriminate amongst various
parametrizations and would enable an extraction of $|V_{ub}|$ from
the experimental branching ratio with
a theoretical error of about 20\%. 
A more effective means of improving the bounds, however, 
would be lattice results over a wider range of $q^2$. This
range will increase as simulations on finer lattices become available.
Of course, a reliable determination of the form factors at or around 
$q^2=0$ would help tremendously. We refrained, in the present
paper, from using sumrule or quark model results around $q^2=0$ because
the systematic errors of these calculations are so different from those
of the lattice. We wanted to keep the results 
model-independent and wished to explore the extent to which
the lattice alone could constrain $\bpi$ decays. In the future, however,
one may want to use the methods developed here to combine results
for the relevant form factors obtained by different methods in different
kinematical regimes.

For the quantities which depend on knowing the $q^2$-dependence
of the form factors close to $q^2_{max}$, such as the $B^*B\pi$ couplings
or partially integrated rates
above the charm production endpoint,
significant improvement in accuracy can be achieved by simply repeating
the simulation which led to the results of \re{jj} with a
sufficient number of light quark masses to enable a numerical chiral
extrapolation of the form factors (see \app{latpar} for details) and
thus provide determinations of the form factors closer to $q^2_{max}$.

We have also performed an analysis of the
QCD corrections--both perturbative and non-perturbative--to the
relevant polarization functions so as to determine the range
of $Q^2$ in which one may trust the QCD evaluation of these functions.
We have found that these corrections are under control
at $Q^2=0$ but not at $Q^2=-16\gev^2$.

Finally, the methods developed here to take into account the
kinematical constraint and errors on the lattice results 
are completely general.
They can be used with sumrule or quark model results to
extend these results to kinematical regimes where these calculations
break down.  Furthermore, our methods are, in
principle, applicable to other processes of great physical interest
such as the rare decay $\bar B\to K^*\gamma$, where a
model-independent guide for extrapolating the lattice results for the
relevant form factor {}from $q^2$ around $q^2_{max}$ to $q^2=0$ 
is urgently needed,
\footnote{Here $q^\mu$ is the momentum of the photon.}
or to the semileptonic decay $\bar
B\to\rho\ell\bar\nu$ which would enable an independent determination
of $|V_{ub}|$. Also, as a by-product of implementing the kinematical
constraint, we have derived a formalism which enables one to constrain
bounds on a form factor with the knowledge that it must lie within an
interval of values at one or more values of $q^2$.  More generally, we
believe that the combination of lattice results with dispersive
techniques will lead to many more interesting results in the future.

\bigskip
\bigskip
\leftline{\large\bf Acknowledgements}

I would like to thank the UKQCD Collaboration and in particular the
authors of \re{jj} for sharing their results with me. I would like to
thank Eduardo de Rafael, Irinel Caprini, Jonathan Flynn, Lawrence
Gibbons, Thomas Mannel, Stephan Narison, Juan Nieves and Jay Watson
for useful discussions.  Eduardo de Rafael, Jonathan Flynn and Juan
Nieves are further thanked for reading the manuscript with care.  I
wish also to thank Alexander Khodjamirian for pointing out a
discrepancy between the normalization convention for $g_{B^*B\pi}$
used in an earlier draft of this paper and the standard convention and
for elaborating on some of the results in Refs.~\cite{VlBK93} and
\cite{VlBB95}.  Finally, I gratefully acknowledge the hospitality of
the Service de Physique Th\'eorique de Saclay and of the Benasque
School for Physics where part of this work was done.

\appendix

\section{Explicit Expressions for the Bounding Functions}
\label{explicit}
Consider the positive semi-definite,
$(L+1)\times(L+1)$ matrix 
\be
M(\vec f) = 
\l(
\begin{array}{ccccc}
\la \phi f|\phi f\ra & \la\phi f|g_{t_1}\ra & \la\phi f|g_{t_2}\ra &\cdots &
\la\phi f|g_{t_L}\ra \\
\la g_{t_1}|\phi f\ra & \la g_{t_1}|g_{t_1}\ra & \la g_{t_1}|g_{t_2}\ra&\cdots
& \la g_{t_1}|g_{t_L}\ra\\
\la g_{t_2}|\phi f\ra & \la g_{t_2}|g_{t_1}\ra & \la g_{t_2}|g_{t_2}\ra&\cdots
& \la g_{t_2}|g_{t_L}\ra\\
\vdots&\vdots&\vdots&\vdots&\vdots\\
\la g_{t_L}|\phi f\ra & \la g_{t_L}|g_{t_1}\ra & \la g_{t_L}|g_{t_2}\ra&\cdots
& \la g_{t_L}|g_{t_L}\ra
\end{array}
\r)
\ ,
\ee
with $\vec f=(f(t_1),\cdots,f(t_L))$ and all $t_i$ distinct.  The
elements of this matrix can be chosen real for $t_1,t_2,\cdots$ in
the interval $(-\infty,t_+)$ in which case the matrix is symmetric.
Then, its determinant is
\be
\det{M(\vec f)} = \la\phi f|\phi f\ra\det{M(\vec f)^{\{(1,1)\}}}-\sum^L_{i,j=1}
a_{ij} \la g_{t_i}|\phi f\ra\la g_{t_j}|\phi f\ra
\label{detm}
\ee
with
\be
a_{ij}=(-1)^{i+j}\det{M(\vec f)^{\{(1,1),(i+1,j+1)\}}}
\label{aij}
\ee
where the matrix $M(\vec f)^{\{(i_1,j_1,),(i_2,j_2),\cdots\}}$ is the matrix
obained by deleting rows $i_1,i_2,\cdots$ and columns $j_1,j_2,\cdots$
{}from $M(\vec f)$. 

Then,
\be
\det{M(\vec f)}\ge 0\Longrightarrow J(Q^2) \det{M(\vec f)^{\{(1,1)\}}}
\ge \sum^L_{i,j=1}
a_{ij} \la g_{t_i}|\phi f\ra\la g_{t_j}|\phi f\ra
\label{poscst}
\ ,\ee
where we have used \eq{jqsqinner} 
and the fact that $\det{M(\vec f)^{\{(1,1)\}}}$
is positive because of the positivity of the inner product.

The positivity of the inner product further implies that:
\footnote{$a_{ii}\ne 0$ and $\det{M(\vec f)^{\{(1,1)\}}}\ne 0$ as long
as the $t_i$, $i=1,\cdots,L$ are distinct.}
\begin{itemize}
\item[1)]
$a_{ii}>0$ for all $i=1,\cdots,L$ so that $\mbox{Tr}\;a>0$;
\item[2)]
$\det{a}=\l(\det{M(\vec f)^{\{(1,1)\}}}\r)^{L-1}>0$.
\end{itemize}
Therefore, the constraint of \eq{poscst} requires $\vec f$ to lie
within the $L$-dimensional volume in $\vec f$-space delimited by the
$L-1$-dimensional ellipsoid centered at the origin and defined by the
equality in \eq{poscst} ($\phi(z(t),Q^2)$ is chosen real and positive
for all $t<t_+$).  One can further show that this ellipsoid is 
circumscribed 
by the box whose sides are given by the hyperplanes which are
solutions to the $L$ equations $\det{M(f_i)}=0$, $i=1,\cdots,L$.  What
this means is that the maximum and minimum values that any of the
$f(t_i)$ can take inside this ellipsoid are given by the unconstrained
bounds of
\eq{fboundsnoinfo} with $t=t_i$.

\bigskip

Now, suppose that we wish to find the bounds that \eq{poscst} imposes
on $f(t_1)$ given $f(t_2),\cdots,f(t_L)$. \eq{poscst}
can be rewritten as
\be
\gamma\ge\alpha \la g_{t_1}|\phi f\ra^2+2\beta\la g_{t_1}|\phi f\ra
\label{quadbnd}
\ee
with
\bea
\alpha &=& a_{11}\ ,\nonumber\\
\beta &=& \sum_{i=2}^L a_{1i} \la g_{t_i}|\phi f\ra\ ,\nonumber\\
\mbox{and}\quad \gamma &=& J(Q^2)\,\det{M(\vec f)^{\{(1,1)\}}}-
\sum^L_{i,j=2}
a_{ij} \la g_{t_i}|\phi f\ra\la g_{t_j}|\phi f\ra
\ .\eea
One can show that the discriminant of the quadratic inequality of \eq{quadbnd}
is
\be
\Delta = \det{M(\vec f)^{\{(1,1)\}}}\,\det{M(\vec f)^{\{(2,2)\}}}
\ ,
\ee
where $M(\vec f)^{\{(2,2)\}}=
M(f(t_2),\cdots,f(t_L))$. Since $\det{M(\vec f)^{\{(1,1)\}}}\ge 0$,
\eq{quadbnd} will have a solution iff $(f(t_2),$ $\cdots,$ $f(t_L))$ 
themselves
satisfy the dispersive constraint $\det{M(f(t_2),\cdots,f(t_L))}\ge 0$.
If this is the case, we find that
\be
F_{lo}(t_1|f(t_2),\cdots,f(t_L))\le f(t_1)\le F_{up}(t_1|f(t_2),\cdots,f(t_L))
\label{appbnds}
\ ,\ee
with
\be
F_{lo/up}(t_1|f(t_2),\cdots,f(t_L))={-\beta\,-{/}+\,\sqrt{\Delta}
\over\alpha}
\label{appbnds1}
\ ,\ee
since $\alpha=a_{11}>0$.
The bounds of \eq{fbounds} in \sec{method} correspond to those of 
\eqs{appbnds}{appbnds1}
with $L=N+1$ and with the replacements $t_1\to t$ and $t_i\to t_{i-1}$,
$i=2,\cdots,N+1$.

\bigskip

Now, to impose the kinematical constraint of \eq{kincst} we need
to know how the bounds $F_{lo}(t|\vec f,x)$ and $F_{up}(t|\vec f,x)$ 
behave as a function of $x$, with 
$\vec f=(f(t_1),\cdots,f(t_N))$ fixed and
$x=f(t_{N+1})$. These bounds are obtained
from the condition $\det{M(f(t),\vec f,x)}\ge 0$. The results of this
appendix imply that the bounds $F_{lo}(t|\vec f,x)$ and 
$F_{up}(t|\vec f,x)$ exist iff 
$\det{M(\vec f,x)}\ge 0$. This in turn means that $x$ is constrained
to the interval $[x_{lo},x_{up}]$ with $x_{lo(up)}=F_{lo(up)}(t_{N+1}|\vec f)$.

Furthermore, because the condition $\det{M(f(t),\vec f,x)}\ge 0$
implies that $(f(t),\vec f,x)$ must lie on an $N+1$-dimensional
ellipsoid, $F_{lo(up)}(t|\vec f,x)$, viewed as a function of $x$ with
$t$ and $\vec f$ fixed, must be the lower (upper) segment of an ellipse
bisected by the line that goes through the points $(x_{lo},
F_{lo}(t|\vec f,x_{lo})$$=$$F_{up}(t|\vec f,x_{lo}))$ and $(x_{up},
F_{lo}(t|\vec f,x_{up})$$=$$F_{up}(t|\vec f,x_{up}))$. Since the
tangents to the ellipse at these points are vertical, 
$F_{up}(t|\vec f,x)$ must increase monotically with $x$ from
$x=x_{lo}$ to $x=x_{max}$ where it reaches its maximum
value and then decrease monotically until $x=x_{up}$
beyond which it is not defined.  Similarly, $F_{lo}(t|\vec f,x)$
decreases monotically with $x$ from $x=x_{lo}$ to
$x=x_{min}$ where it reaches its minimum value and then increases
monotically until $x=x_{up}$ beyond which it is not
defined.
\footnote{$x_{min}$ and $x_{max}$ may equal $x_{lo}$ or $x_{up}$.}
Moreover, one can show that
$F_{up}(t|\vec f,x_{max})=F_{up}(t|\vec f)$, i.e. the bounds
one would obtain without any constraints on $x=f(t_{N+1})$. Simlarly,
$F_{lo}(t|\vec f,x_{min})=F_{lo}(t|\vec f)$.

\section{The Functions $\chi_{T,L}(Q^2)$ in QCD}
\label{pert}
To quantify the $\ord{\a_s}$ corrections to the polarization function,
$\Pi^{\mu\nu}(q)$, we use the results of \re{genthesis,LJRe81}.
Because the mass of the $b$ quark, $m_b$, appears already at 1-loop
order, the relative size of this leading contribution and the
$\ord{\a_s}$ correction will depend on the renormalization scheme used
for this mass. The calculations of \re{genthesis,LJRe81} were peformed
using an on-shell scheme. To test the reliability of these
calculations as $Q^2$ is changed, we perform them in two additional
schemes. Thus, the three schemes are:
\begin{itemize}
\item[1)] pole: 
\begin{itemize}
\item[$\bullet$]
$m_b = m_b^{pole}$ 
\item[$\bullet$]
$\a_s(\mu)$ in $\msbar$-scheme 
\item[$\bullet$]
$\mu=\sqrt{Q^2+m_b^2}$
\end{itemize}
\item[2)] $\msbar$:
\begin{itemize}
\item[$\bullet$]
$m_b = \overline{m_b}(\mu)$ with $m_b^{pole}=
m_b\l(1+\frac{\a_s(\mu)}{\pi}\phi(\mu)\r)$
and $\phi(\mu)=\ln{
\frac{\mu^2}{m_b^2}}+\frac{4}{3}$
\item[$\bullet$]
$\a_s(\mu)$ in $\msbar$-scheme
\item[$\bullet$]
$\mu=\sqrt{Q^2+m_b^2}$
\end{itemize}
\item[3)] euclidean (Landau gauge):
\begin{itemize}
\item[$\bullet$]
$m_b = m_b^{euc}(\mu)$ with $m_b^{pole}=
m_b\l(1+\frac{\a_s(\mu)}{\pi}\phi(\mu)\r)$
and $\phi(\mu)=-\frac{\mu^2+m_b^{pole}}{\mu^2}\ln{
\frac{\mu^2+m_b^{pole}}{(m_b^{pole})^2}}$
\item[$\bullet$]
$\a_s(\mu)$ in $\msbar$-scheme
\item[$\bullet$]
$\mu=\sqrt{Q^2+(m_b^{pole})^2}$.
\end{itemize}
\end{itemize}
$\mu=\sqrt{Q^2+m_b^2}$ is a natural scale for this process and, obviously, 
$\phi\equiv 0$ in the pole scheme.

The
non-perturbative corrections, which are due to interactions of the $b$
and $\bar u$ quarks with quark and gluons condensates, can be
evaluated with the standard diagrammatic techniques of \re{svz} or
with background field methods as in \re{gennonpert}. Using
the diagrammatic techniques, we have checked the results of
\re{LJRe81,gennonpert} for condensates of dimensions 3 and 4. Combining all
of these results and keeping contributions of condensates with
dimension less or equal to 4, we find (setting $m_u=0$)
\bea
\chi_L(Q^2) & = & \frac{1}{\pi m_b^2}\int_0^1
dx\,\frac{\l(m_b^2/x\r)\,\im\Pi_L^{pert}(x)}{\l(1+(Q^2/m_b^2) x\r)^2}
\nonumber\\
&& + \frac{\bar m_b(1\gev)\la\bar uu\ra_{1\gev}}{(Q^2+m_b^2)^2} 
+ \frac{1}{(Q^2+m_b^2)^2}\la\frac{\a_s}{12\pi}G^2\ra
\ ,
\label{chilqcd}
\eea
with
\bea
\l(m_b^2/x\r)\ \im\Pi_L^{pert}(x) 
&=& \frac{3 m_b^2}{8\pi} (1-x)^2\l\{ 1+\frac{4\a_s}{3\pi}
\l(\frac{9}{4}+2 l(x)+\ln{x}\ln{1-x}\r.\r.\nonumber\\
&& \l.\l.+\l(\frac{5}{2}-x-\frac{1}{1-x}\r)
\ln{x}+\l(x-\frac{5}{2}\r)\ln{1-x}+3\phi(\mu)\frac{Q^2x}{m_b^2+Q^2x}\r)\r\}
\ ,
\label{impil}
\eea
and
\bea
\chi_T(Q^2) & = & \frac{1}{\pi m_b^2}\int_0^1
dx\,\frac{\im\Pi_T^{pert}(x)}{\l(1+(Q^2/m_b^2) x\r)^3}
\nonumber\\
&& -\frac{\bar m_b(1\gev) \la\bar uu\ra_{1\gev}}{(Q^2+m_b^2)^3}
-\frac{1}{(Q^2+m_b^2)^3}\la\frac{\a_s}{12\pi}G^2\ra 
\ ,
\label{chitqcd}
\eea
with
\bea
\im \Pi_T^{pert}(x) &=& \frac{1}{8\pi} (1-x)^2\l\{(2+x)+\frac{4\a_s}{3\pi}
\l[(2+x)\l(\frac{13}{4}+2 l(x)+\ln{x}\ln{1-x}+\frac{3}{2}
\ln{\frac{x}{1-x}}\r.\r.\r.\nonumber\\
&&\l. -\ln{1-x}-x\ln{\frac{x}{1-x}}-\frac{x}{1-x}
\ln{x}\r)\nonumber\\
&&-\l((3+x)(1-x)\ln{\frac{x}{1-x}}+
\frac{2x}{(1-x)^2}\ln{x}+5+2x+\frac{2x}{1-x}\r)\nonumber\\
&&\l.\l.+\frac{3}{2}\phi(\mu)\frac{2Q^2x-m_b^2}{m_b^2+Q^2x}\r]\r\}
\ .
\label{impit}
\eea
In \eqs{impil}{impit}, $l(x)$ is a dilogarithm. 

We have evaluated the expressions of \eqs{chilqcd}{chitqcd} for two
values of $Q^2$ to investigate the range of validity of the QCD
calculation. One might like to take $Q^2$ close to the resonance
region because the strength of the bounds may increase as $Q^2$
approaches this region as suggested by Eqs.~(\ref{disprelS}),
(\ref{disprelV}), (\ref{chil}) and (\ref{chit}). We have chosen
$Q^2=0$, which is the value traditionally used when dealing with
current correlators involving a heavy quark, and $Q^2=-16\gev^2$ as
was suggested in \re{boyd}.  For the masses, we use the results
of \re{narmb} ($\bar{m}_b(\bar{m}_b)=4.3\gev$) and for the condensates:
\bea
\la\bar uu\ra_{1\gev} = (-0.24\gev)^3&&\nonumber\\
\la\frac{\a_s}{\pi} G^2\ra = 0.02(1)\gev^4&&
\ ,
\label{condens}
\eea
where the quark condensate is taken {}from~\cite{PaB93} 
and the gluon condensate, {}from~\cite{nargg}. 
Our results for the different contributions
to $\chi_L(Q^2)$ and $\chi_T(Q^2)$ are summarized in \tab{xltab}
and \tab{xttab}, respectively.
\begin{table}[tb]
\begin{center}
\begin{tabular}{|c|c|c|c|c|c|c|} \hline
&\multicolumn{6}{c|}{$\chi_L(Q^2)$}\\ 
\hline 
$Q^2$ & \multicolumn{3}{c|}{$0\gev^2$} & \multicolumn{3}{c|}{$-16\gev^2$}
\\ \hline
scheme & pole & $\overline{\mbox{MS}}$ & Euclid. & pole & $\overline{\mbox{MS}}$ & Euclid. \\ \hline
1-loop & $1.3\expo{-2}$ & $1.3\expo{-2}$ & $1.3\expo{-2}$ 
& $2.1\expo{-2}$ & $2.1\expo{-2}$ & $2.6\expo{-2}$ \\ \hline
$\ord{\a_s}$/1-loop & 16\% & 17\% & 16\% & 35\% & 32\% & 12\%\\ \hline
total pert. & $1.5\expo{-2}$ & $1.5\expo{-2}$ & $1.5\expo{-2}$ 
& $2.8\expo{-2}$ & $2.8\expo{-2}$ & $2.9\expo{-2}$ \\ \hline
$m_u\la\bar uu\ra$/1-loop & -1\% & -2\% & -2\% & -11\% & -10\% & -66\% 
\\ \hline
$\la\a_s G^2\ra$/1-loop & $0.03\%$ & $0.04\%$ & $0.04\%$ 
& $0.2\%$ 
& $0.2\%$ & $1.2\%$ \\ \hline
total & $1.5\expo{-2}$ & $1.5\expo{-2}$ & $1.4\expo{-2}$ 
& $2.6\expo{-2}$ & $2.6\expo{-2}$ & $1.2\expo{-2}$\\ \hline
\end{tabular}
\caption{Perturbative and non-perturbative contributions to the 
subtracted polarization function 
$\chi_L(Q^2)$ at two values of $Q^2$.
\label{xltab}}
\end{center}
\end{table}
\begin{table}[tb]
\begin{center}
\begin{tabular}{|c|c|c|c|c|c|c|} \hline
&\multicolumn{6}{c|}{$\chi_T(Q^2)$}\\ 
\hline 
$Q^2$ & \multicolumn{3}{c|}{$0\gev^2$} & \multicolumn{3}{c|}{$-16\gev^2$}
\\ \hline
scheme & pole & $\overline{\mbox{MS}}$ & Euclid. & pole & $\overline{\mbox{MS}}$ & Euclid. \\ \hline
1-loop & $4.2\expo{-4}$ & $5.1\expo{-4}$ & $5.1\expo{-4}$ 
& $1.0\expo{-3}$ & $1.0\expo{-3}$ & $2.1\expo{-3}$ \\ \hline
$\ord{\a_s}$/1-loop & 25\% & 7\% & 6\% & 44\% & 38\% & -40\%\\ \hline
total pert. & $5.3\expo{-4}$ & $5.5\expo{-4}$ & $5.4\expo{-4}$ 
& $1.5\expo{-3}$ & $1.4\expo{-3}$ & $1.3\expo{-3}$ \\ \hline
$m_u\la\bar uu\ra$/1-loop & 2\% & 3\% & 3\% & 33\% & 30\% & 352\% \\ \hline
$\la\a_s G^2\ra$/1-loop & $-0.03\%$ & $-0.05\%$ & $-0.05\%$ 
& $-0.6\%$ 
& $-0.5\%$ & $-6.2\%$ \\ \hline
total & $5.3\expo{-4}$ & $5.6\expo{-4}$ & $5.6\expo{-4}$ 
& $1.8\expo{-3}$ & $1.7\expo{-3}$ & $8.5\expo{-3}$\\ \hline
\end{tabular}
\caption{Perturbative and non-perturbative contributions to the 
subtracted polarization function 
$\chi_T(Q^2)$ at two values of $Q^2$.
\label{xttab}}
\end{center}
\end{table}

At $Q^2=0$, the perturbative and non-perturbative corrections to the
one-loop results for $\chi_L(Q^2)$ and $\chi_T(Q^2)$ are under control,
especially when the perturbative $\msbar$ and euclidean schemes are used.
Moreover, there is very little scheme dependence in the total perturbative
contribution and actual values of $\chi_L(0)$ and $\chi_T(0)$.
Thus, the QCD calculation appears to be reliable at $Q^2=0$.
For $Q^2=-16\gev^2$, even though the scheme dependence of the total
perturbative contributions to $\chi_L(Q^2)$ and $\chi_T(Q^2)$ is small,
the size of the $\ord{\a_s}$ corrections are large and it is
somewhat doubtful that the perturbative series will converge. Furthermore,
the quark condensate contributions become quite large and even
blow up in the euclidean scheme. So even though the values of
$\chi_L(Q^2)$ and $\chi_T(Q^2)$ are nearly equal in the pole and
$\msbar$ schemes, one cannot really trust the various expansions
at $Q^2=-16\gev^2$. 
We therefore restrict ourselves to $Q^2=0$ throughout the paper
and use the $\msbar$ results which appear to be the most convergent
and which give the loosest constraint on the polarization functions
at $Q^2=0$.

\section{Lattice Parameters and Discussion of Systematic Errors}
\label{latpar}
The results for the form factors $f^+$ and $f^0$ presented in \tab{latres}
were obtained by the UKQCD Collaboration {}from 60 
quenched $SU(3)$
gauge configurations on a $24^3\times 48$ lattice at $\beta=6.2$~\cite{jj}.
The corresponding inverse lattice spacing as determined
{}from the string tension is $a^{-1}=2.73(5)\gev$~\cite{qlhms}.
Heavy and light-quark propagators were obtained using an $\ord{a}$-improved
Sheikholeslami-Wohlert action~\cite{sw} and $\ord{a}$-improved
currents~\cite{heatlie}. 
\begin{table}[tb]
\begin{center}
\begin{tabular}{|c||c|c|} \hline
$q^2\ (\gev^2)$ & $f^+(q^2)$ & $f^0(q^2)$\\ \hline 
14.9%3 
& $0.85%\er{6}{6}
\pm 0.20$ & $0.46%\er{3}{3}
\pm 0.10$\\ 
17.2%3 
& $1.10%\er{5}{5}
\pm 0.27$ & $0.49%\er{2}{2}
\pm 0.10$\\ 
20.0%3 
& $1.72%\err{22}{22}
\pm 0.50$ & $0.56%\er{3}{3}
\pm 0.12$\\ \hline
\end{tabular}
\caption{Lattice results for $f^+(q^2)$ and $f^0(q^2)$ at three values 
of $q^2$. Please see text for details on the treatment of systematic errors.
\label{latres}}
\end{center}
\end{table}

Because cutoff effects would be too large, the $b$-quark cannot
be simulated directly. Therefore, the three-point functions used
to determine the form factors were calculated at four values
of the heavy-quark mass straddling that of the charm quark
\footnote{The corresponding hopping parameters are 0.121, 0.125,
0.129, 0.133.} and the corresponding results for $f^+$ and $f^0$ were
extrapolated in heavy-quark mass to $m_B$ according to
\eqs{hqsfo}{hqsfp} allowing for linear corrections in $1/m_B$ and
including in the errors the variations due to possible $1/m_B^2$
corrections~\cite{jj}. Of the six extrapolated points in \re{jj}, we
have kept only those for which the corresponding recoil $\w$ is
completely independent of the heavy-quark mass so as to limit the
introduction of possible systematic effects.
\footnote{The value of $f^+(q^2)$ at $q^2=20.0\gev^2$ was obtained
from a heavy-quark mass extrapolation--identical to the one of 
\protect\re{jj}--of the four values of $f^+(q^2_{max})$ 
corresponding to the four heavy-quark masses already described. These
four values of $f^+(q^2_{max})$ were obtained, in turn, from a
``pole/dipole'' fit to the results for $f^+(q^2)$ and $f^0(q^2)$ at
each individual heavy-quark mass value~\protect\cite{drb}.}

To keep volume errors under control, one cannot work with arbitrarily
light quarks. The strategy, then, is to perform the calculation for
several values of light-quark mass around the mass of the strange and
then extrapolate the results to the up and down.  However, the limited
number of light-quark mass values available to the authors of
\re{jj} made it impossible for them to perform 
a reliable chiral extrapolation.
Therefore, the results presented in \tab{latres} were obtained with
light-quark masses slightly larger than that of the strange
($\k_{light}=0.14144$) and light-flavor symmetry is assumed for both the
active and spectator light quarks. 

Dependence of $f^+$ and $f^0$ on light spectator-quark mass should
be quite small.  This is corroborated by the results
of~\cite{dtok} for semileptonic $D\to K,\pi$ decays where the form
factors display no statistically significant
variation as the mass of the spectator quark is changed {}from
slightly above $m_{strange}$ to zero, while holding $q^2$ fixed.
Since it is these very same data--supplemented with results at
different values of the charm quark mass--which were used in \re{jj}
to obtain the lattice results quoted here, we expect 
spectator-quark flavor symmetry to be reasonable. In fact, similar
spectator-quark flavor symmetry is observed in the form factors which
govern \bks\ decays~\cite{bks} and in those which govern semileptonic
$\brhogen$ decays~\cite{JoFN9506} where it is as small as 5\%.  
We will assume here that uncertainties associated
with light-spectator-quark flavor symmetry are 10\% on both form factors for
all $q^2$.
We do not expect much of a $q^2$-dependence in the corrections to
spectator-quark flavor symmetry for $f^+$ since 
in a pole dominance scenario, which suits the data quite well, 
the mass of 
the relevant pole does not depend on the spectator's flavor.
For $f^0$, which has a less pronounced $q^2$-dependence, the situation
should be even better.

For the values of $q^2$ considered here (see \tab{latres}), we also expect
dependence on the active light-quark mass to be reasonable.  All
recent lattice simulations of semileptonic $D$ decays indicate that the
ratio $f^+_{D\to \pi}(0)/f^+_{D\to K}(0)$ is consistent with 1
within errors. For instance, the authors of \re{dtok} find $f^+_{D\to
K}(0)/f^+_{D\to\pi}(0)=1.09\pm 0.09$.  Furthermore, 
the author of the sumrule calculation of \re{su3nar} finds that
active light-quark flavor symmetry at $q^2=0$ is even better for semileptonic 
$B$ decays: $f^+_{D\to K}(0)/f^+_{D\to \pi}(0)=1.10\pm0.01$ 
versus $f^+_{\bar B\to
K}(0)/f^+_{\bar B\to \pi}(0)=1.01\pm0.02$.
And because of the kinematical constraint of
\eq{kincst}, these conclusions also hold for $f^0(0)$.
To get a measure of the
size of the uncertainties associated with assuming
this flavor symmetry for fixed $q^2$ away from 0, we suppose
pole dominance for $f^+(q^2)$. Such behavior is consistent with the
$q^2$-dependence of the lattice results as well as with our bounds.
Then, assuming active-quark flavor symmetry at $q^2=0$, 
we find that $f^+_{\bar B\to
\pi}(q^2)/ f^+_{latt}(q^2)\simeq
(1-q^2/m^2_{B_s^*})/(1-q^2/m^2_{B^*})$ ranges {}from 1.04 to 1.09
for the $q^2$ in \tab{latres} and for $m_{B_s^*}=5.42\gev$~\cite{pdg}.
Thus, adding 4\% to 9\% errors to the uncertainty on active-quark 
flavor symmetry 
at $q^2=0$ should
give a reasonable estimate of the possible errors on $f^+$ for 
$q^2\ne 0$.
For $f^0$, which has a milder $q^2$-dependence than that
given by pole dominance, we expect that the dependence on $q^2$ of 
uncertainties due to the assumption
of active-quark flavor
symmetry to be mild: they
should be of the same order as those on $f^0(0)$.
Now, to accomodate a possible, slight $q^2$-dependence of these uncertainties,
as well as possible deviations from our pole parametrization for the 
$q^2$-dependence of $f^+$ and violations 
of light, active-quark flavor symmetry 
at $q^2=0$, we will assume that this symmetry holds within 10\%
at $q^2=0$.
Thus, we add in quadrature to the
uncertainties in $f^+$, errors ranging {}from 14 to 19\% and to
those in $f^0$, 10\% errors.

The error associated with ignoring internal quark loops (i.e.
quenching) is difficult to quantify. However, quenched calculations of
form factors for semileptonic $D\to K,\,K^*$ decays (see \cite{dtok}
and references therein) give results in agreement with world average
experimental values, while quenched calculations of the differential
decay rate for semileptonic $\bar B\to D$ and $\bar B\to D^*$ decays 
agree very well with
experimental data~\cite{wisprd}.  This gives us confidence that errors
due to quenching for processes involving heavy-light hadrons are rather
small.
Moreover, the errors due to the uncertainty in the scale discussed in
the next paragraph should take some of the uncertainties associated
with quenching into account since quenching is in large part responsible
for this scale uncertainty. Nevertheless, to account for 
quenching errors which are not taken into account by
the scale uncertainty we add a conservative 10\% error to the form factors.

In obtaining the bounds, we have assumed that the inverse lattice
spacing is $a^{-1}=2.7 \gev$, as is given by the $\rho$-meson mass
($a^{-1}(m_\rho)=2.7(1)\gev$~\cite{strange}) or the string tension
($a^{-1}(\sigma)=2.73(1)\gev$~\cite{qlhms}) and in accordance with the
choice made in~\cite{jj}. Other physical quantities lead to slightly
different values for the inverse lattice spacing, typically in the
range $2.5\sim 2.9\gev$ for spectral quantities~\cite{strange}. This
uncertainty in the determination of the scale obviously leads to
uncertainties in the lattice results. These errors should be
relatively small, here, since the form factors are dimensionless
quantities.  There are, however, two places where the scale appears.
The first is in expressing the heavy-light pseudoscalar masses in
physical units for the heavy-quark-mass extrapolations of the form
factors. The second is in the determination of $q^2$ in physical
units, but only in so far as the scale is used to determine the mass of the
final state meson (the ``pion''), since we obtain $q^2$ {}from the
dimensionless quantity $\w$ through $q^2=m_B^2+m_\pi^2-2m_Bm_\pi\w$,
using the experimental value for $m_B$. We have quantified these
errors by using the heavy-quark-mass dependence of the form factors
given in \re{jj} and the $q^2$-dependences obtained in
\sec{comparison}. We find that they are less than about 5\% for
both form factors.

Finally, there are cutoff corrections of order $\ord{\a_s am_Q}$ and
$\ord{(am_Q)^2}$, where $m_Q$ is the mass of the heavy quark in
the initial meson. For the heavy-quark masses and action
used in \re{jj}, these lattice artefacts induce errors in the
form factors which range {}from 5 to 10\%
as indicated in the study of \re{wisprd}. A systematic
study of similar effects is also given in~\cite{zvzavlad}. Assuming,
for simplicity, that these errors grow linearly with $m_Q$, we
have estimated the size of the error induced in the value of the
form factors at the scale of the $b$ quark. We have found them
to be about 10\%.

To summarize, the errors on the lattice results for $f^+$ and $f^0$
given in \tab{latres} are the quadratic sum of the statistical
uncertainties given in~\cite{jj}; of a 10\% systematic error
associated with possible corrections to spectator-quark flavor
symmetry; of an error ranging {}from 14\% to 19\%, depending on $q^2$,
to account for possible deviations {}from light-active-quark flavor
symmetry in $f^+$ and a corresponding 10\% error in $f^0$; of a 10\%
error associated with quenching; of a 5\% error to take into account
uncertainties associated with the determination of the lattice scale;
and finally, of a 10\% error to account for possible cutoff effects.
With these systematic errors (and in some cases even
without) the lattice results of \protect\tab{latres} are compatible
with the very few chirally extrapolated points of Refs.
\protect\cite{ELC94} and
\protect\cite{apebpi} within one standard deviation or less.
Finally, because our errors include both statistical and systematic
uncertainties, all of the results in the present paper incorporate a
full treatment of statistical and systematic errors.


\begin{thebibliography}{99}

\bibitem{bpicleo}
R.~Ammar {\it et al.} (CLEO Collaboration), CLEO CONF 95-09, EPS0165,
July 1995.
%``Semi-Leptonic B Decays at CLEO'',
%talk given the {\it $XXX^{th}$ Rencontres de Moriond: QCD and High
%Energy Hadronic Interactions}, Les Arcs, March 19-25, 1995.

\bibitem{pdg}
Particle Data Group, \prd{50} (1994) 1173.

\bibitem{NaIS89}
N.~Isgur, D.~Scora, B.~Grinstein and M.~B.~Wise, \prd{39} (1989) 799.

\bibitem{WSB85}
M.~Wirbel, B.~Stech and M.~Bauer, \zpc{29} (1985) 637.

\bibitem{KS88}
J.G.~ K\"orner and G.A.~Schuler, \zpc{38} (1988) 511; 
{\bf 41} (1989) 690(E).

\bibitem{PaOX95}
P.J.~O'Donnell, Q.P.~Xu and H.K.K.~Tung, \prd{52} (1995) 3966.

\bibitem{DaSI95}
D.~Scora and N.~Isgur, \prd{52} (1995) 2783.
%CEBAF-TH-94-14, hep-ph/9503486.

\bibitem{ChCHZ96}
C-Y.~Cheung, C-W.~Hwang and W-M.~Zhang, IP-ASTP-16-95 and hep-ph/9602309.

\bibitem{IGNS96}
I.L.~Grach, I.M.~Narodetskii and S.~Simula, INFN-ISS 96/4 and hep-ph/9605349.

\bibitem{CeDP88}
C.A.~Dominguez and N.~Paver, \zpc{41} (1988) 217.

\bibitem{AO89}
A.A.~Ovchinnikov, \plb{229} (1989) 127.

\bibitem{StN92}
S.~Narison, \plb{283} (1992) 384

\bibitem{StN95}
S.~Narison, \plb{345} (1995) 166.

\bibitem{PaB93}
P.~Ball, \prd{48} (1993) 3190.

\bibitem{VlBK93}
V.M. Belyaev, A. Khodjamirian and R.~R\"uckl, \zpc{60} (1993) 349.

\bibitem{VlBB95}
V.M. Belyaev, V.M.~Braun, 
A. Khodjamirian and R.~R\"uckl, \prd{51} (1995) 6177.

\bibitem{AlKR95}
A. Khodjamirian and R.~R\"uckl, {\it talk presented by R.~R\"uckl at
``Beauty 95'', Oxford, July 1995, to appear in the proceedings},
MPI-PhT/95-97, LMU 18/95 and hep-ph/9510294.

\bibitem{NaIW90}
N.~Isgur and M.B.~Wise, \prd{41} (1990) 151.

\bibitem{NaIW90.2}
N.~Isgur and M.B.~Wise, \prd{42} (1990) 2388.

\bibitem{GuBD92}
G.~Burdman and J.F.~Donoghue, \plb{280} (1992) 287.

\bibitem{MaW92}
M.B.~Wise, \prd{45} (1992) 2188.

\bibitem{LiW92}
L.~Wolfenstein, \plb{291} (1992) 177.

\bibitem{GuBLNN94}
G.~Burdman, Z.~Ligeti, M.~Neubert and Y.~Nir, \prd{49} (1994) 2331.

\bibitem{akhoury}
R.~Akhoury, G.~Sterman and Y.-P.~Yao, \prd{50} (1994) 358.

\bibitem{ELC94}
ELC Collaboration, As.~Abada {\it et al.}, \npb{416} (1994) 675.

\bibitem{apebpi}
APE Collaboration, C.~R.~Allton {\it et al.}, \plb{345} (1995) 513;
V.~Lubicz, private communication.

\bibitem{jj}
UKQCD Collaboration, D.~R.~Burford {\it et al.},
\npb{447} (1995) 425. 
% Glasgow GUTPA-95-2-1,
%FERMILAB-PUB-95-023-T, Southampton SHEP 95-09 and
%hep-lat/9503002.

\bibitem{drb}
D.~R.~Burford, private communication.

\bibitem{machet}
C.~Bourrely, B.~Machet and E.~de Rafael, \npb{189} (1981) 157.

\bibitem{boyd}
C.~G.~Boyd, B.~Grinstein and R.~F.~Lebed, \prl{74} (1995) 4603.
%UCSD preprint
%UCSD/PTH 94-27 and hep-ph/9412234.

\bibitem{CGBGL96}
C.~G.~Boyd, B.~Grinstein and R.~F.~Lebed, \npb{461} (1996) 493.

\bibitem{IrCN96}
I.~Caprini and M.~Neubert, CERN-TH/95-255 and hep-ph/9603414.

\bibitem{CGBL96}
C.~G.~Boyd and R.~F.~Lebed, UCSD/PTH 95-23 and hep-ph/9512363.

\bibitem{qhldc}
UKQCD Collaboration, R.~M.~Baxter {\it et al.}, \prd{49} (1994) 1594.

%\bibitem{abada}
%As. Abada, LPTHE Orsay-94/79 and hep-ph/9409338.

\bibitem{BeGM94}
B.~Grinstein and P.~Mende, \npb{425} (1994) 451.

%\bibitem{BeGM93}
%B.~Grinstein and P.~Mende, \plb{299} (1993) 127.

\bibitem{josep}
E.~de Rafael and J.~Taron, \prd{50} (1994) 373.

\bibitem{okubo}
S.~Okubo and I-Fu Shih, \prd{4} (1971) 2020.

\bibitem{eric}
L.~Lellouch, E.~Sather and L.~Randall, \npb{405} (1993) 55.

\bibitem{PiCD94}
R.~Colangelo, F.~De Fazio and G.~Nardulli, \plb{334} (1994) 175.

\bibitem{HaDN95}
H.~G.~Dosch and S.~Narison, \plb{368} (1996) 163.
%Montpellier preprint PM-95-41 and
%hep-ph/9510212.

\bibitem{zak}
L.~Lellouch in the proceedings of the $XXXIV^{th}$ Cracow School of
Theorectical Physics, Acta Physica Polonica {\bf B25} (1994) 1679.

%\bibitem{genpert}
%S.~C.~Generalis, J.~Phys.~{\bf G16} (1990) 785.

\bibitem{genthesis}
S.~C.~Generalis, Open Univserity preprint OUT-4102-13.

\bibitem{LJRe81}
L.J. Reinders, S. Yazaki and H.R. Rubinstein, \plb{103} (1981) 63.

\bibitem{svz}
M.~A.~Shifman, A.~I.~Vainshtein and V.~I.~Zakharov, \npb{147} (1979) 385.

\bibitem{gennonpert}
S.~C.~Generalis, J.~Phys.~{\bf G16} (1990) 367.


\bibitem{narmb}
S.~Narison, Acta Physica Polonica {\bf B26} (1995) 687.

\bibitem{CaBNRY81}
C.~Becchi, S.~Narison, E.~de Rafael and F.~J.~Yndur\' ain, \zpc{8} (1981) 335.


\bibitem{nargg}
E.~Braaten, S.~Narison and A.~Pich, \npb{373} (1992) 581.

\bibitem{qlhms}
UKQCD Collaboration, C.~Allton {\it et al.}, \npb{407} (1993) 331.

\bibitem{sw}
B.~Sheikholeslami and R.~Wohlert, \npb{259} (1985) 572.

\bibitem{heatlie}
G.~Heatlie, G.~Martinelli, C.~Pittori, G.~C.~Rossi and
C.~T.~Sachrajda, \npb{352} (1991) 266.

\bibitem{dtok}
UKQCD Collaboration, K.~C.~Bowler {\it et al.}, \prd{51} (1995) 4905.

\bibitem{bks}
UKQCD Collaboration, K.~C.~Bowler {\it et al.}, \prd{51} (1995) 4955.
%Edinburgh-94-554 and hep-lat/9407013.

\bibitem{su3nar}
S.~Narsion, \plb{337} (1994) 163.

\bibitem{wisprd}
UKQCD Collaboration, K.~C.~Bowler {\it et al.}, \prd{52} (1995) 5067;
%Edinburgh-93/525,
%Marseille CPT-95/P.3179, Southampton SHEP 95-06 and hep-ph/9504231; 
L.~Lellouch (for the
UKQCD Collaboration) in the proceedings of the $30^{th}$ {\it
Recontres de Moriond: QCD and High Energy Hadronic Interactions},
Meribel les Allues, France, 19-25 March 1995, Marseille CPT-95/P.3196 
and hep-ph/9505423.

\bibitem{strange}
UKQCD Collaboration, C.~Allton {\it et al.}, \prd{49} (1994) 474.

\bibitem{zvzavlad}
M.~Crisafulli, V.~Lubicz, G.~Martinelli and A.~Vladikas, Nucl.~Phys.~{\bf B}
(Proc.~Suppl.) {\bf 42} (1995) 400.

\bibitem{JoFN9506}
UKQCD Collaboration, J.~M.~Flynn {\it et al.}, \npb{461} (1996) 327.
%and hep-ph/9506398.

\bibitem{IKr96}
I.J. Kroll, FERMILAB-CONF-96-032, to appear in {\it Proceedings
of the XVII International Symposium on Lepton-Photon Interactions},
Beijing, China, 10-15 August 1995.

\end{thebibliography}
\end{document}